\documentclass[useAMS,usenatbib]{mn2e}

\usepackage{amssymb,graphicx,multirow,txfonts}

\newcommand{\etal}{et~al.}
\newcommand{\PVdblt}{{\rm P}\kern 0.1em{\sc v}~$\lambda\lambda 1117, 1128$}
\newcommand{\CaIIdblt}{{\rm Ca}\kern 0.1em{\sc ii}~$\lambda\lambda 3934, 3969$}
\newcommand{\AlIIIdblt}{{\rm Al}\kern 0.1em{\sc iv}~$\lambda\lambda 1855, 1863$}
\newcommand{\CIVdblt}{{\rm C}\kern 0.1em{\sc iv}~$\lambda\lambda 1548, 1550$}
\newcommand{\MgIIdblt}{{\rm Mg}\kern 0.1em{\sc ii}~$\lambda\lambda 2796, 2803$}
\newcommand{\NVdblt}{{\rm N}\kern 0.1em{\sc v}~$\lambda\lambda 1238, 1242$}  
\newcommand{\SVIdblt}{{\rm S}\kern 0.1em{\sc vi}~$\lambda\lambda 933, 944$} 
\newcommand{\OVIdblt}{{\rm O}\kern 0.1em{\sc vi}~$\lambda\lambda 1031, 1037$} 
\newcommand{\SiIIdblt}{{\rm Si}\kern 0.1em{\sc ii}~$\lambda\lambda 1190, 1193$} 
\newcommand{\SiIVdblt}{{\rm Si}\kern 0.1em{\sc iv}~$\lambda\lambda 1393, 1402$} 
\newcommand{\AlI}{\hbox{{\rm Al}\kern 0.1em{\sc i}}}
\newcommand{\AlII}{\hbox{{\rm Al}\kern 0.1em{\sc ii}}}
\newcommand{\AlIII}{{\hbox{\rm Al}\kern 0.1em{\sc iii}}}
\newcommand{\CaII}{\hbox{{\rm Ca}\kern 0.1em{\sc ii}}}
\newcommand{\CII}{\hbox{{\rm C}\kern 0.1em{\sc ii}}}
\newcommand{\CIIe}{\hbox{{\rm C$^{\ast}$}\kern 0.1em{\sc ii}}}
\newcommand{\CIII}{\hbox{{\rm C}\kern 0.1em{\sc iii}}}
\newcommand{\CIV}{\hbox{{\rm C}\kern 0.1em{\sc iv}}}
\newcommand{\CV}{\hbox{{\rm C}\kern 0.1em{\sc v}}}
\newcommand{\HI}{\hbox{{\rm H}\kern 0.1em{\sc i}}}
\newcommand{\HII}{\hbox{{\rm H}\kern 0.1em{\sc ii}}}
\newcommand{\Lya}{\hbox{{\rm Ly}\kern 0.1em$\alpha$}}
\newcommand{\Lyb}{\hbox{{\rm Ly}\kern 0.1em$\beta$}}
\newcommand{\Lyg}{\hbox{{\rm Ly}\kern 0.1em$\gamma$}}
\newcommand{\Lyd}{\hbox{{\rm Ly}\kern 0.1em$\delta$}}
\newcommand{\Lye}{\hbox{{\rm Ly}\kern 0.1em$\epsilon$}}
\newcommand{\Lyf}{\hbox{{\rm Ly}\kern 0.1em$\zeta$}}
\newcommand{\HeI}{\hbox{{\rm He}\kern 0.1em{\sc i}}}
\newcommand{\HeII}{\hbox{{\rm He}\kern 0.1em{\sc ii}}}
\newcommand{\FeI}{\hbox{{\rm Fe}\kern 0.1em{\sc i}}}
\newcommand{\FeII}{\hbox{{\rm Fe}\kern 0.1em{\sc ii}}}
\newcommand{\FeIII}{\hbox{{\rm Fe}\kern 0.1em{\sc iii}}}
\newcommand{\MnII}{\hbox{{\rm Mn}\kern 0.1em{\sc ii}}}
\newcommand{\MgI}{\hbox{{\rm Mg}\kern 0.1em{\sc i}}}
\newcommand{\MgII}{\hbox{{\rm Mg}\kern 0.1em{\sc ii}}}
\newcommand{\MgIII}{\hbox{{\rm Mg}\kern 0.1em{\sc iii}}}
\newcommand{\NI}{\hbox{{\rm N}\kern 0.1em{\sc i}}}
\newcommand{\NII}{\hbox{{\rm N}\kern 0.1em{\sc ii}}}
\newcommand{\NIII}{\hbox{{\rm N}\kern 0.1em{\sc iii}}}
\newcommand{\NV}{\hbox{{\rm N}\kern 0.1em{\sc v}}}
\newcommand{\OVI}{\hbox{{\rm O}\kern 0.1em{\sc vi}}}
\newcommand{\OI}{\hbox{{\rm O}\kern 0.1em{\sc i}}}
\newcommand{\OII}{\hbox{[{\rm O}\kern 0.1em{\sc ii}]}}
\newcommand{\OIII}{\hbox{[{\rm O}\kern 0.1em{\sc iii}]}}
\newcommand{\SI}{{\rm S}\kern 0.1em{\sc i}}
\newcommand{\SIV}{{\rm S}\kern 0.1em{\sc iv}}
\newcommand{\SVI}{{\rm S}\kern 0.1em{\sc vi}}
\newcommand{\SiI}{\hbox{{\rm Si}\kern 0.1em{\sc i}}}
\newcommand{\SiII}{\hbox{{\rm Si}\kern 0.1em{\sc ii}}}
\newcommand{\SiIII}{\hbox{{\rm Si}\kern 0.1em{\sc iii}}}
\newcommand{\SiIV}{\hbox{{\rm Si}\kern 0.1em{\sc iv}}}
\newcommand{\SII}{\hbox{{\rm S}\kern 0.1em{\sc ii}}}
\newcommand{\SIII}{\hbox{{\rm S}\kern 0.1em{\sc iii}}}
\newcommand{\NaI}{\hbox{{\rm Na}\kern 0.1em{\sc i}}}
\newcommand{\TiII}{\hbox{{\rm Ti}\kern 0.1em{\sc ii}}}
\newcommand{\ZnII}{\hbox{{\rm Zn}\kern 0.1em{\sc ii}}}
\newcommand{\CrII}{\hbox{{\rm Cr}\kern 0.1em{\sc ii}}}
\newcommand{\kms}{\hbox{km~s$^{-1}$}}
\newcommand{\cmsq}{\hbox{cm$^{-2}$}}

\def\aj{{AJ}}                   
\def\araa{{ARA\&A}}             
\def\apj{{ApJ}}                 
\def\apjl{{ApJ}}                
\def\apjs{{ApJS}}

\def\mnras{{MNRAS}}

\def\pasp{{PASP}}

\def\nat{{Nature}}

\title[Cold accretion at z=0.7]{Discovery of multi-phase cold accretion in a massive galaxy at z=0.7}

\author[G. G. Kacprzak et al.]{Glenn G. Kacprzak,$^{1,2}$\thanks{gkacprzak@astro.swin.edu.au} 
Christopher W. Churchill,$^{3,4}$
Charles C. Steidel,$^{5}$
\newauthor Lee R. Spitler,$^{1}$
and Jon A. Holtzman,$^{3}$ 
\\
$^{1}$ Centre for Astrophysics and Supercomputing, Swinburne University of Technology, PO Box 218, Victoria 3122, Australia\\
$^{2}$ Australian Research Council Super Science Fellow\\
$^{3}$ Department of Astronomy, New Mexico State University, Las Cruces, NM 88003\\
$^{4}$ Visiting Professor, Swinburne University of Technology, Victoria 3122, Australia\\
$^{5}$ California Institute of Technology, MS 105-24, Pasadena, CA 91125, USA}

\begin{document}
\date{Accepted August 16 2012}

\pagerange{\pageref{firstpage}--\pageref{lastpage}} \pubyear{2010}

\maketitle

\label{firstpage}

\begin{abstract}
We present detailed photo+collisional ionization models and kinematic
models of the multi-phase absorbing gas, detected within the {\it
HST}/COS, {\it HST}/STIS, and Keck/HIRES spectra of the background
quasar TON 153, at 104~kpc along the projected minor axis of a
star-forming spiral galaxy ($z=0.6610$).  Complementary $g'r'i'Ks$
photometry and stellar population models indicate that the host galaxy
is dominated by a $\sim4$ Gyr stellar population with slightly greater
than solar metallicity and has an estimated log$M_{\ast}=11$ and a
log$M_{\rm vir}=13$.  Photoionization models of the low ionization
absorption, ({\MgI}, {\SiII}, {\MgII} and {\CIII}) which trace the
bulk of the hydrogen, constrain the multi-component gas to be cold
(log$T=3.8-5.2$) and metal poor ($-1.68\leq [X/H] \leq -1.64$).  A
lagging halo model reproduces the low ionization absorption
kinematics, suggesting gas coupled to the disk angular momentum,
consistent with cold accretion mode material in simulations.  The
{\CIV} and {\OVI} absorption is best modeled in a separate
collisionally ionized metal-poor ($-2.50\leq [X/H] \leq -1.93$) warm
phase with log$T=5.3$.  Although their kinematics are consistent with
a wind model, given the $2-2.5$~dex difference between the galaxy
stellar metallicity and the absorption metallicity indicates the gas
cannot arise from galactic winds. We discuss and conclude that
although the quasar sight-line passes along the galaxy minor axis at
projected distance of 0.3 virial radii, well inside its virial shock
radius, the combination of the relative kinematics, temperatures, and
relative metallicities indicated that the multi-phase absorbing gas
arises from cold accretion around this massive galaxy.  Our results
appear to contradict recent interpretations that absorption probing
the projected minor axis of a galaxy is sampling winds.


\end{abstract}


\section{Introduction}

Over the last decade, simulations have shown that galaxy evolution is
highly dependent on gas accretion occurring via two modes: hot and
cold accretion. Current cosmological simulations demonstrate that the
majority of gas accreted at early epochs onto galaxies occurs via the
cold mode, which has temperatures of $T\sim10^4-10^5$~K and
metallicities of $Z\lesssim 0.01Z_{\odot}$. Cold-mode gas is
preferentially accreted along cosmic filaments/streams and have high
densities and low cooling times providing a large supply of gas
penetrating through hot halos surrounding galaxies
\citep{keres05,dekel06,ocvirk08,keres09,brooks09,
  dekel09,ceverino10,stewart11a,stewart11b,freeke11a,freeke11b,faucher-giguere11}.
It is expected that cold accretion should comprise no more than ~7\%
of the total {\HI} mass density at $z\sim 1$ \citep{kacprzak11c}.

It is further expected that cold accretion truncates when the host
galaxy mass exceeds $\sim10^{12}$~M$_{\odot}$, since infalling gas
becomes shock heated to the halo viral temperature ($\sim 10^6$) and
is predicted to dramatically reduce the cold accretion cross-section
to a tiny fraction
\citep[e.g.,][]{dekel06,keres05,ocvirk08,dekel09,keres09,stewart11a,
brooks09,freeke11a,freeke11b} of the observed halo gas cross-section
\citep{kacprzak08,chen10a}.  However, it is expected that these dense
filaments can still survive within hot halos and could provide an
efficient means of feeding massive galaxies with pristine gas
\citep[e.g.,][]{keres05}.

The study of absorbing foreground gas detected in background quasar
spectra allows us to probe these otherwise unobservable comic
filaments and outflows. {\MgII} absorption is ideal for detecting cold
mode and hot mode accretion, wind outflows, etc., since it probes gas
with a large range of neutral hydrogen column densities, $10^{16}
\lesssim \hbox{N(\HI)} \lesssim 10^{22}$~{\cmsq}
\citep{archiveI,weakII} with gas temperature of around
30,000$-$40,000~K and average total hydrogen densities of
$\sim0.1$~atoms~cm$^{-3}$ \citep{cvc03,ding05}. It has also been
thoroughly demonstrated that {\MgII} absorption is produced within
gaseous halos surrounding galaxies and is not produced within the
intergalactic medium (IGM) \citep[see][]{cwc-china}.

Over the last decade, strong {\MgII} absorption has also been observed
to directly trace 100$-$1000~{\kms} galactic-scale outflows
\citep{tremonti07,weiner09,martin09,rubin10,coil11,martin12} that
extend out to at least 50~kpc along the galaxy minor axis
\citep{bordoloi11, bouche11,kacprzak12}. Galactic winds have been
observed over a large range of redshifts and detected using a range of
ions \citep[see][ and references therein]{steidel10}. Correlations
between galaxy colors and star formation rates with {\MgII} equivalent
widths also indirectly suggest that absorption is produced in outflows
\citep{zibetti07,noterdaeme10,nestor11}.

However, {\MgII} has been observed infalling \citep{martin12} into
highly inclined galaxies with velocities of 100$-$200~{\kms}
\citep{rubin11}. This is consistent with \citet{kacprzak11b} who
showed that absorption strength is correlated with the orientation of
the galaxy major axis, implying that a significant fraction of weaker
{\MgII} absorption systems are likely accreting toward the galaxy via
cold flows. The bimodal azimuthal angle distribution of quasar
sight-lines around {\MgII} absorption selected galaxies also suggests
that infall occurs along the projected galaxy major axis
\citep{bouche11,kacprzak12}.  These cold-flow streams likely produce a
circumgalactic co-rotating gas component that is predominately
infalling towards the galaxy and, in absorption, these structures are
expected to have $\sim 100$~{\kms} velocity offsets relative to the
host galaxy {\it and} in the same direction of galaxy rotation
\citep{stewart11b}. These models are consistent with previous
observations of \citet{steidel02} and \citet{kacprzak10a} that show
{\MgII} absorption residing fully to one side of the galaxy systemic
velocity and usually aligned with expected galaxy rotation direction,
with the absorption essentially mimicking the extension of the galaxy
rotation curve out into the halo. We expect low ionization states,
such as {\MgI}, {\MgII}, {\SiII}, {\CII} and {\CIII} to be ideal for
tracing cold mode accretion given metallicity, temperatures and
densities expected.

A reliable means of determining the origins of the absorbing gas is to
obtain both the host-galaxy and absorption-line metallicity.
Absorption-line metallicities for a handful of systems have been
determined to range between [M/H] $<-1.8$ to $-1$ while existing near
sub-L$^{\star}$ galaxies that have nearly solar metallicities
\citep{zonak04,chen05, tripp05, cooksey08, kacprzak10b, ribaudo11,
thom11}. It is postulated that these extremely low metallicity
absorption systems are likely accreting onto their host galaxies and
possibly trace cold mode accretion, which is still expected for these
sub-L$^{\star}$ galaxies. In the rare case where absorption-line
metallicities are larger than the host galaxy is suggestive that the
absorption is probing winds \citep{peroux11}.

Here we target a particular galaxy that has {\MgII} absorption
consistent with disk-like kinematics, possibly tracing cold
accretion. The absorption also contains a separate warm phase as
indicated by separate strong {\CIV} absorption that does not coincide
with {\MgII}. We have obtained supplementary {\it HST}/COS data in
order to determine the physical properties of the gas.  In this paper,
we perform kinematic and photo+collisional ionization models of
multi-phase absorbing gas obtained from {\it HST}/COS, {\it HST}/STIS,
and Keck/HIRES, which is associated with star-forming spiral galaxy at
$z=0.6610$.  In \S~\ref{sec:data} we describe our targeted galaxy
and our data. We discuss our absorption-line analysis in
\S~\ref{sec:anal}. In \S~\ref{sec:galresults} we describe the host
galaxy properties determined from broad-band photometry and stellar
population models.  In \S~\ref{sec:absresults} we describe the results
of our kinematics and photo+collisional ionization models and the
physical properties of the absorbing gas.  In \S~\ref{sec:dis}, we
discuss the possible origins of the absorption and our
concluding remarks are in \S~\ref{sec:conclusion}. Throughout we adopt
an H$_{\rm 0}=70$~\kms Mpc$^{-1}$, $\Omega_{\rm M}=0.3$,
$\Omega_{\Lambda}=0.7$ cosmology.

\section{Target Field and Observations}
\label{sec:data}

\begin{figure*}
\includegraphics[angle=0,scale=0.44]{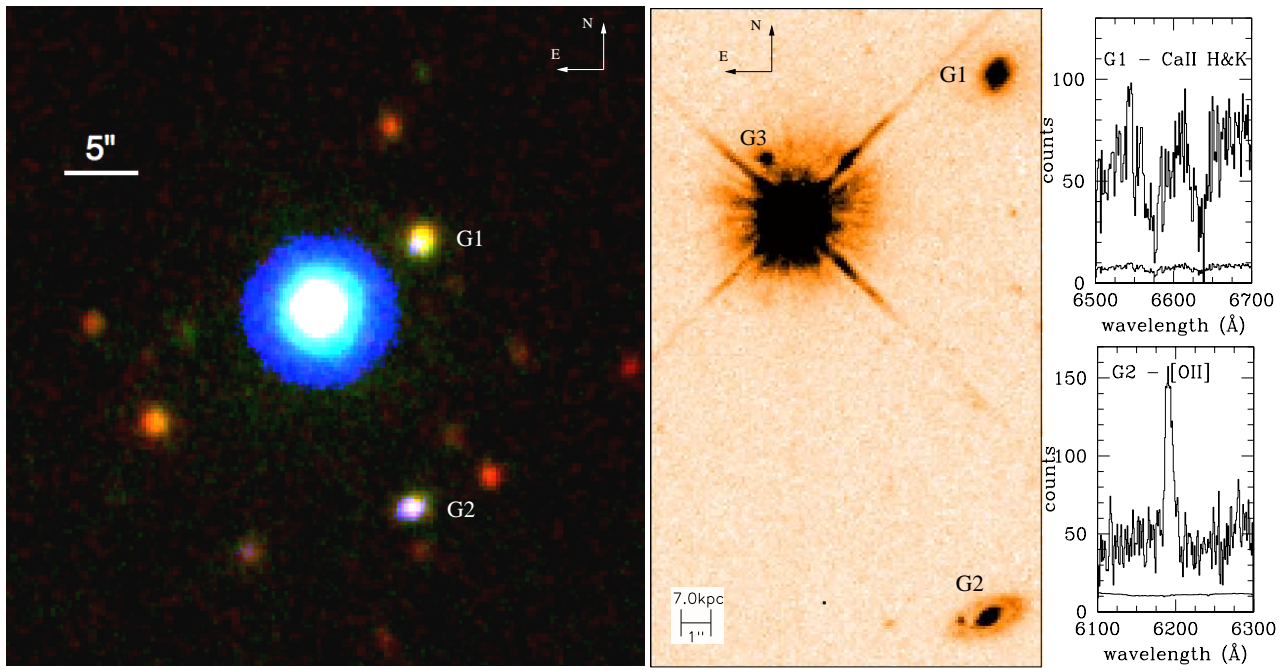}
\caption{--- (left) Ground-based $g'-$ (blue), $i'-$ (green) and $Ks-$
  band (red) color composite centered on the quasar. Note that G1 is
  redder than G2 \citep[see][]{churchill12b}. Also note that the
  surrounding galaxies in the field are extremely red compared to both
  $z\sim 0.6$ galaxies, indicating that they are at much higher
  redshifts and not associated with the absorption. (middle) A
  $13''\times22''$ {\it HST}/WFPC-2 F702W image of the quasar field
  TON 153. The quasar is the brightest object in the northeast corner
  of the image.  G2 is associated with a variety of metal-lines shown
  in Figure~\ref{fig:trans}.  A possible object seen within
  $\sim$1$''$ northeast of the quasar (G3) has not been successfully
  identified spectroscopically since the quasar is bright. Note the
  quasar sight-line passes near the minor axis of the moderately
  inclined disk of G2. --- (top right) A Keck/LRIS spectrum of G1
  containing {\CaII} H \& K confirms the redshift of G1 to be
  $z_{gal}=0.6719$.  --- (bottom left) A Keck/LRIS spectrum of G2
  containing {\OII} emission confirms the redshift of G2 to be
  $z_{gal}=0.6610$.}
\label{fig:qsofield}
\end{figure*}

TON 153, also known as Q1317$+$227, is a bright (V=16.0 mag) quasar at
$z_{em}=$1.017. Inspection of a low resolution quasar spectrum
revealed two {\MgII} absorption systems at $z_{abs}=0.29$ and
$z_{abs}= 0.66$ \citep{ss92}. Following a spectroscopic survey of
galaxies in close angular proximity to the quasar sight-line,
\citet{steidel02} discovered galaxies G1 and G2 shown in
Figure~\ref{fig:qsofield}. The quiescent early-type galaxy G1 has a
redshift of $z_{gal}=0.6719$ and the star-forming disk galaxy G2 has a
redshift of $z_{gal}=0.6610$. \citet{churchill07} demonstrated that G1
was associated with a {\Lya} complex that did not have any observable
metals even though it resides at $D=58.1$~kpc; well within the 100~kpc
where metals are expected
\citep[e.g.,][]{chen01,kacprzak08,chen10a,tumlinson11}. In a companion
paper \citep{churchill12b}, we further discuss G1 and its associated
absorption lines. The galaxy G2 is at $D=103.9$~kpc and is associated
with extensive metal-line/LLS absorption at $z_{abs} = 0.6601$
\citep{ss92,bachall93,bahcall96,archiveI,ding05,churchill07} and is
the focus of this paper.

\begin{figure*}
\includegraphics[angle=0,scale=1.85]{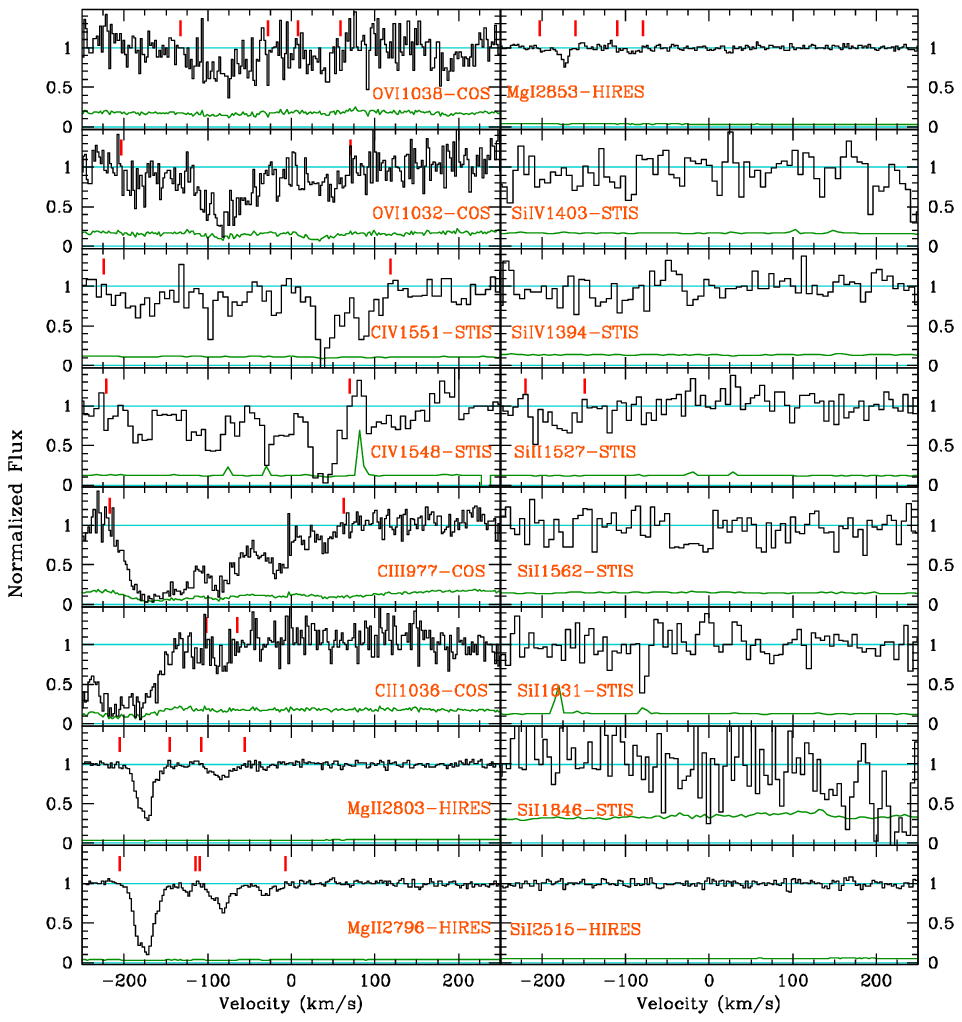}
\caption{--- Metal absorption-lines associated with the $z=0.6610$
galaxy G2 obtained from {\it HST}/COS, {\it HST}/STIS and
Keck/HIRES. In each panel the transition and instrument are
labeled. The absorption-line data (black) are relative to the systemic
velocity of G2. The regions highlighted by the tick-marks (red)
indicate the detected absorption features and the location of the
tick-marks are defined to be where the equivalent width per resolution
element recovers to a $1~\sigma$ detection threshold. The sigma
spectrum is shown below the data (green).}
\label{fig:trans}
\end{figure*}

\subsection{HST Imaging}

In Figure~\ref{fig:qsofield} we present a 4700 second {\it
  HST}/WFPC--2 F702W image (PID 5984; PI Steidel) that was reduced
using the WFPC-2 Associations Science Products Pipeline
(WASPP\footnote{http://archive.stsci.edu/hst/wfpc2/pipeline.html}). Note
that the F702W filter provides a bandpass similar to a rest-frame
Johnson B-band filter for galaxies at $z\sim0.6$.  Galaxy magnitudes
and luminosities were obtained from \citet{churchill12a} and are based
on the AB system. The GIM2D software \citep{simard02} was used to
obtain quantified galaxy morphological parameters that were published
in \citet{kacprzak11b} and \citet{churchill12a}.

\subsection{Ground-based Imaging}

To further constrain the properties of G2, we analyzed multi-band
$g'r'i'$ and $Ks$ imaging.  The $g'r'i'$-bands were obtained using
SPIcam CCD imager on the APO 3.5 m telescope. SPIcam has a field of
view of $4.78 \times 4.78'$ and a spatial resolution of
$0.14''$/pixel. Our observations were taken with on-chip binning of $2
\times 2$ which provides a plate scale of 0.28$''$/pixel. The images
were taken as part of a large survey program and were observed over 4
nights between March 2006 and March 2007 providing total exposure
times in the $g'r'i'$ of 5190, 4630 and 4350 seconds, respectively,
with typical seeing of $1.1-1.6''$.

Multiple frames were taken in each filter and each frame was
individually reduced using standard IDL and IRAF\footnote{IRAF is
written and supported by the IRAF programming group at the National
Optical Astronomy Observatories (NOAO) in Tucson, Arizona. NOAO is
operated by the Association of Universities for Research in Astronomy
(AURA), Inc.\ under cooperative agreement with the National Science
Foundation.} packages. Pixel-to-pixel variations were removed using a
combination of dome and twilight sky flat fields. Due to PSF/seeing
variations over the long exposures, cosmic rays were removed from each
frame separately. The SPIcam pixels are sufficiently small that
interpolation errors do not lead to significant photometric
uncertainties.  The astrometry was calibrated by matching field stars
from each frame to USNO A2.0 catalog stars. 

The photometric zeropoints were established using a number of stars
from the SDSS catalog.  Since the APO $g'r'i'$ filter/detector
combination does not exactly match that of the SDSS survey, there can
exist color terms between the two systems. We derived these color
terms by analyzing the SDSS stellar photometry from over 30 fields and
produced a single photometric solution.  We adopt the color term
derived from the large set of observations and the average corrections
are of the order of 0.1 magnitudes.

Deep near-infrared (Ks) band images were obtained in 1994 February 24
using the Kitt Peak 4-m Mayall telescope and the IRIM NICMOS 3 $256
\times 256$ camera with 0.6$''$/pixel. These images were part of a
more extensive {\MgII} galaxy survey of \citet{sdp94}. The images were
reduced using the contributed IRAF package DIMSUM\footnote{{\it
http://iraf.noao.edu/iraf/ftp/contrib/dimsumV3/.} DIMSUM was
contributed by P. Eisenhardt, M. Dickinson, S. A. Stanford, \&
F. Valdez.}. The photometric zeropoints were established using a
number of stars from the 2MASS point-source catalog
\citep{skrutskie06}.

In Figure~\ref{fig:qsofield} we present a $g'i'Ks$ color composite
image centered on TON 153 with galaxies G1 and G2 labeled.  See
companion paper \citet{churchill12b} for further discussions of the
properties of G1. Note that all other objects near the quasar
sight-line are red compared to the two $z\sim 0.65$ galaxies,
indicating that all the other objects are likely at much higher
redshifts and that G1 and G2 are likely isolated objects.
\citet{churchill12b} noted that the ROSAT X-ray luminosity limit for
this field is four orders of magnitude less than what is expect for
cluster centers and is consistent with the expected luminosity for
early-type galaxy halos at $z=0.67$. In addition, the absorption
metallicities determined by \citet{churchill12b} for G1, and those
found here, are 1$-$2dex lower than expected for inter-cluster
medium. Thus, it is likely that the absorption does not arise in a
cluster environment.

The galaxy G3 is located within $\sim$1$''$ northeast of the quasar
sight-line and has not been successfully identified spectroscopically
since the quasar is bright. It is possible that this object is
responsible for either the absorption associated with G1, G2, the
$z=0.29$ absorption system, absorption at the quasar redshift, or
other metal-lines identified in this sight-line. If G3 is at $z=0.67$,
than \citet{churchill12b} estimates its mass to be a factor of ten
less than the mass of G2 and would have a similar metallicity to G2
according to mass-gas metallicity relations
\citep[e.g.,][]{savaglio05}. Thus, if G3 is at the same redshift of
G2, it could be considered as a satellite of G2 and would occupy the
same gaseous and dark matter halo.

Photometry for calibration and science was extracted using {\tt
  SExtractor} \citep{bertin96} using the {\tt MAGAUTO} measurements.
Corrections for Galactic dust extinction were applied to the galaxies
using the dust maps of \citet{schlegel98}. We obtained final dust-,
color- and seeing-corrected AB magnitudes for G1 of $m_{g'}=23.43\pm
0.03$, $m_{r'}=22.17\pm0.03$, $m_{i'}=21.01\pm 0.03$ and
$m_{Ks'}=19.4\pm 0.1$ \citep{churchill12b} and for G2 of
$m_{g'}=23.23\pm 0.03$, $m_{r'}=22.32\pm0.03$, $m_{i'}=21.41\pm 0.03$
and $m_{Ks'}=19.9\pm 0.1$.


%
%
%



\subsection{Galaxy Spectroscopy}

The galaxies shown in Figure~\ref{fig:qsofield} were spectroscopically
identified by \citet{steidel02} and their spectra were first presented
in \citet{churchill07}. A possible object seen within $\sim$1$''$ from
the quasar seen in the image has not been successfully identified
spectroscopically since the quasar is bright.  The details of the
Keck/LRIS spectroscopic observations can be found in \citet{steidel02}
and \citet{churchill07,churchill12b}. The spectra are both vacuum and
heliocentric velocity corrected. Galaxy G1, identified by {\CaII} H \&
K absorption yields a $z_{gal}=0.6719$, is associated with broad
{\Lya} complex that spans 1400~{\kms} and yet contains only very weak
metals lines \citep{churchill07, churchill12b}. Galaxy G2 was
identified by a {\OII} emission line, placing it at
$z_{gal}=0.6610$. This galaxy is associated with extensive
metal-line/LLS absorption at $z_{abs} = 0.6601$
\citep{bachall93,bahcall96,archiveI,ding05,churchill07} and is the
focus of this paper. We discuss G2's associated metals-lines in the
next sections.

\subsection{Quasar Spectroscopy}

Some of the $z_{abs} = 0.6601$ absorption properties were measured
from a 3600 second Keck/HIRES ($R\sim 45,000$) exposure observed in
1995 January. The details of the observation and data reduction are
described in \citet{cv01}. In addition, a 12,000 second {\it HST}/STIS
E230M ($R\sim 30,000$) exposure was obtained (PID 8672; PI Churchill)
and the data reduction details are found in \citet{ding05}.

Recent {\it HST}/COS observations of the quasar ($R\sim 18,000$) were
obtained using the FUV G160M grating and the NUV G185M grating (PID
11667; PI Churchill).  The FUV observations were centered at
1600~{\AA} and took place on 25 June 2010 with an exposure time of
12,580 seconds.  The NUV observations consisted of two exposures that
occurred on 26 May 2010 and were optimally co-added.  The first 5420
second NUV exposure was centered at 1921~{\AA} and the 4970 second NUV
exposure was centered at 1941~{\AA}. The COS spectra were reduced
using the standard {\it HST} IRAF pipeline.  All spectra are both
vacuum and heliocentric velocity corrected.

It is important to note that the wavelength solutions across these
multiple instruments are consistent at the sub-pixel level. This was
verified across all spectrographs by centroiding common ionization
absorption lines from the data presented here and the data from
\citet{churchill12b}.

Analysis of the absorption profiles was performed using interactive
software \citep[see][]{weakI,archiveI,cv01} for local continuum
fitting, objective feature identification, and measuring absorption
properties. Velocity widths and equivalent widths of the absorption
systems are measured between the pixels where the equivalent width per
resolution element recovers to the $1~\sigma$ detection threshold
\citep{weakI}.

\begin{figure}
\includegraphics[angle=0,scale=0.43]{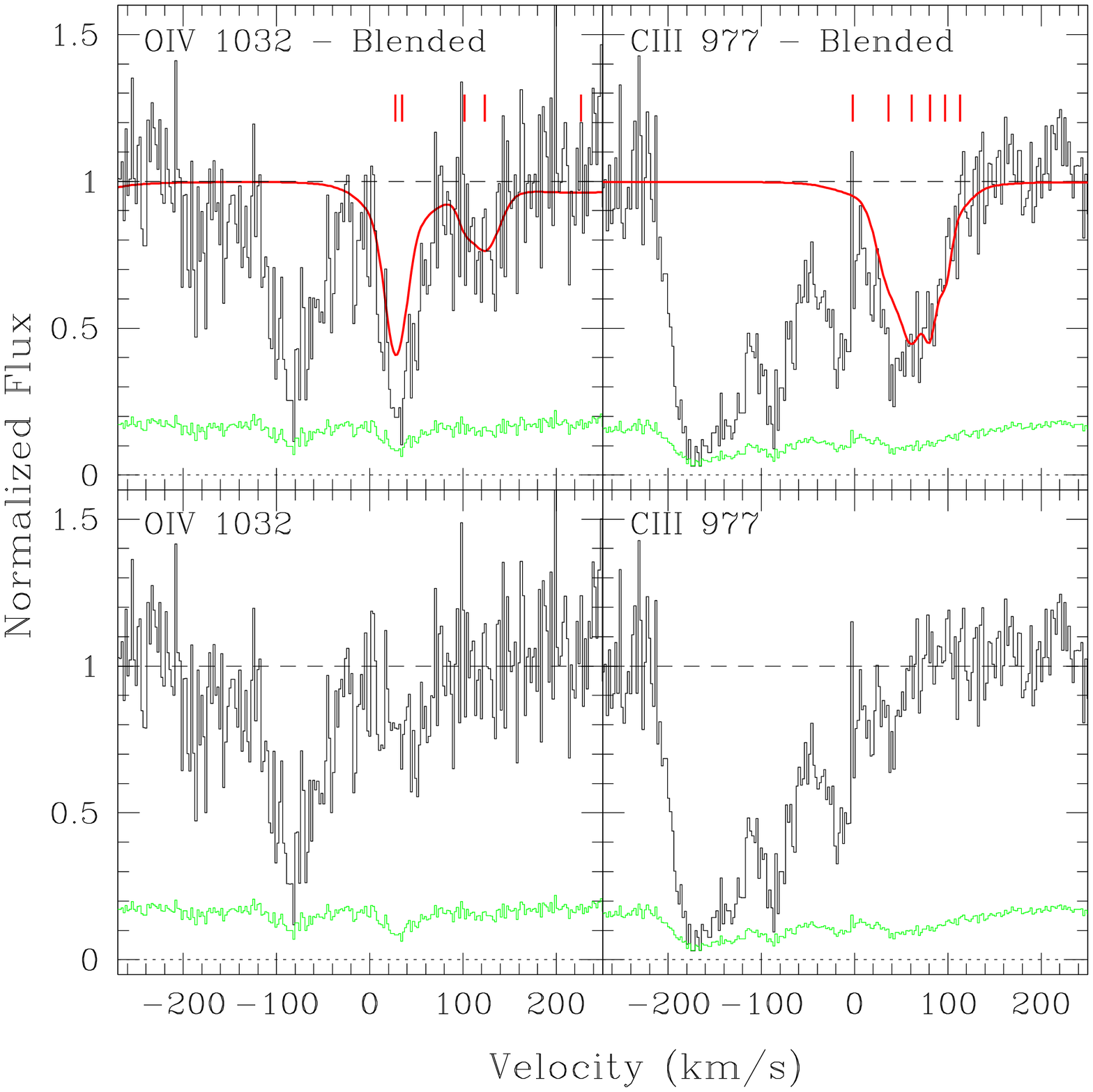}
\caption{--- (top-left) COS/HST {\OVI}~$\lambda$1032 transition
  associated with G2 is blended with {\Lyb} from the $z=$0.67 G1
  {\Lya} complex \citep{churchill07,churchill12b}.  The {\Lyb} fit
  (red) was constrained by simultaneous Voigt profile fits to the
  {\Lya}, {\Lyg}, {\Lye}, and {\Lyf}.  --- (top-right) COS/HST
  {\CIII}~$\lambda$977 transition associated with G2 is blended with a
  {\Lyg} from the $z=$0.67 G1 {\Lya} complex
  \citep{churchill07,churchill12b}.  The Voigt profile fit was
  constrained by the {\Lya}, {\Lyb}, {\Lye}, and {\Lyf} which produce
  the {\Lyg} fist shown (red).  --- (bottom-left) The deblended
  {\OVI}1032 line associated with G2.  --- (bottom-right) The
  deblended {\CIII}~$\lambda$977 line associated with G2.}
\label{fig:blend}
\end{figure}

\section{Quasar Absorption-line Analysis}
\label{sec:anal}

In Figure~\ref{fig:trans} we present the absorption-line data and in
Table~\ref{tab:abs} we present the measured equivalent widths and
column densities. We discuss below how the column densities and/or
limits were computed for each transition. We further discuss how we
account for higher redshift hydrogen-line blends that contaminate the
{\OVI}~$\lambda$1032 and {\CIII}~$\lambda$977 transitions (see
Figure~\ref{fig:blend}). 

The {\HI} column density was adopted from \citet{churchill07},
obtained by simultaneously fitting the Ly$\alpha$, Ly$\beta$, and
Lyman break obtained from a FOS spectrum, was determined to be
$\log[N({\HI})]=18.3\pm0.3$. This is consistent with the results of
\citet{rao06}, who fit only the Ly$\alpha$ line, and determined the
column density to be $18.57\pm 0.02$. If we were to use the Rao et al
value, our derived metallicities would decrease be 0.27 dex.

We also note upon analyzing the {\CIVdblt} doublet, we have deemed the
reddest component seen in the {\CIV}~$\lambda$1551 transition, at
roughly 80~{\kms}, to be real and has been included in our
analysis. This feature does not appear in the {\CIV}~$\lambda$1548
line since this region is contaminated by poor sky subtraction.  This
component was omitted by the analysis of \citet{ding05} but included
here.

\begin{figure}
\includegraphics[angle=0,scale=0.53]{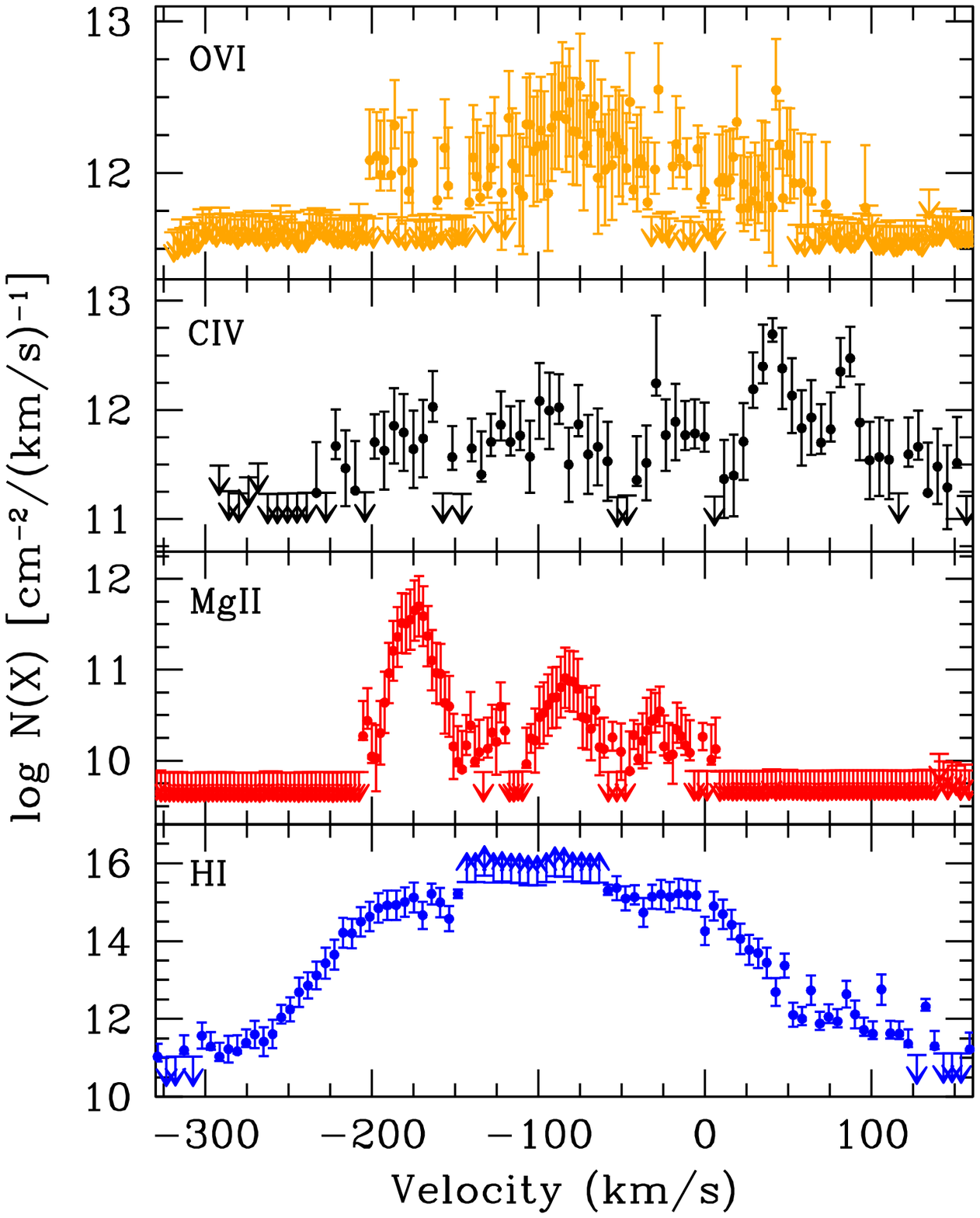}
\caption{--- The doublet optimally combined apparent optical depth
distribution for {\OVI} (orange), {\CIV} (black) and {\MgII} (red).
The {\HI} (blue) optical depth distribution was produced by combining
the entire Lyman series. The total column densities are integrated
over the velocity bins. Note that the {\HI} is still saturated in the
center of the profile. Also note the velocity structure difference
between the profiles and that the majority of the hydrogen is at
velocities consistent with the {\MgII} whereas the majority of the
{\CIV} has velocities consistent with a small fraction of the total
{\HI}. }
\label{fig:AOD}
\end{figure}

\begin{table*}
\begin{center}
  \caption{Absorption Properties$^a$}
  \vspace{-0.5em}
\label{tab:abs}
{\footnotesize\begin{tabular}{lcccrrrr}\hline
Ion                    & Instrument/  &         $W_r$        &    log$N(X)$            &$W_r^{blue}\phantom{00}$&  $^b$log$N(X)^{blue}$     &$W_r^{red}\phantom{00}$& $^b$log$N(X)^{red}$ \\
                       &  Telescope   &         (\AA)        &                         & (\AA)$^b\phantom{00}$  &                       & (\AA)$^b\phantom{00}$ &            \\\hline
{\MgII}~$\lambda$2796  &HIRES/Keck    &  0.348$\pm$0.007     &                         &  0.348$\pm$0.007     &                         &$<$0.0049        &                    \\
{\MgII}~$\lambda$2803  &HIRES/Keck    &  0.198$\pm$0.006     &                         &  0.198$\pm$0.006     &                         & $<$0.0060       &                    \\
{\bf MgII}            &              &                      & 13.11$^{+0.07}_{-0.07}$ &                      & 13.11$^{+0.07}_{-0.07}$ &                 &    $\leq$11.06       \\[1.0ex]
{\MgI}~$\lambda$2853   &HIRES/Keck    &  0.039$\pm$0.005     &                         &  0.039$\pm$0.005     &                         & $<$0.0049       &                    \\
{\bf MgI}              &              &                      & 11.54$^{+0.06}_{-0.05}$ &                      & 11.54$^{+0.05}_{-0.05}$ &                 &   $\leq$10.58       \\[1.0ex]
{\SiI}~$\lambda$2515   &HIRES/Keck    & $<$0.0066            &                         & $<$0.0066            &                         &  $<$0.0066      &                      \\
{\SiI}~$\lambda$1846   &STIS/{\it HST}& $<$0.053             &                         & $<$0.053             &                         &  $<$0.053       &                     \\
{\SiI}~$\lambda$1631   &STIS/{\it HST}& $<$0.019             &                         & $<$0.019             &                         &  $<$0.019       &                    \\
{\SiI}~$\lambda$1562   &STIS/{\it HST}& $<$0.023             &                         & $<$0.023             &                         &  $<$0.023       &                    \\
{\bf SiI}              &              &                      &   $\leq$11.8            &                      &   $\leq$11.8            &                 & $\leq$11.8        \\[1.0ex]
{\SiII}~$\lambda$1527  &STIS/{\it HST}&0.067$\pm$0.013       &                         &0.067$\pm$0.013       &                         &  $<$0.018       &                    \\
{\bf SiII}             &              &                      &  13.16$^{+0.11}_{-0.08}$&                      &  13.16$^{+0.11}_{-0.08}$&                 & $\leq$12.6        \\[1.0ex]
{\SiIV}~$\lambda$1394  &STIS/{\it HST}& $<$0.019             &                         & $<$0.019             &                         &  $<$0.019       &                    \\
{\SiIV}~$\lambda$1403  &STIS/{\it HST}& $<$0.023             &                         & $<$0.023             &                         &  $<$0.023       &                     \\
{\bf SiIV}             &              &                      &   $\leq$12.4            &                      &   $\leq$12.4            &                 & $\leq$12.4       \\[1.0ex]
{\CII}~$\lambda$1036   &COS/{\it HST} &$>$0.018$\pm$0.005$^c$&                         &$>$0.018$\pm$0.005$^c$&                         &$<$0.012         &                  \\
{\bf CII}              &              &                      &$>13.39^{+0.09}_{-0.07}$ &                      &$>13.39^{+0.09}_{-0.07}$ &                 &  $\leq$13.1      \\[1.0ex]
{\CIII}~$\lambda$977   &COS/{\it HST} & 0.467$\pm$0.008      &                         & 0.431$\pm$0.007      &                         &0.036$\pm$0.004  &                   \\
{\bf CIII}             &              &                      & 14.20$^{+0.03}_{-0.02}$ &                      & 14.18$^{+0.03}_{-0.03}$ &                 &12.88$^{+0.05}_{-0.05}$\\[1.0ex]
{\CIV}~$\lambda$1548   &STIS/{\it HST}&$>$0.508$\pm$0.027$^d$&                         &  0.315$\pm$0.024     &                         &$>$0.193$\pm$0.031$^d$&                \\
{\CIV}~$\lambda$1551   &STIS/{\it HST}& 0.417$\pm$0.026      &                         &  0.189$\pm$0.021     &                         & 0.228$\pm$0.018 &                     \\         
{\bf CIV}              &              &                      &  14.41$^{+0.06}_{-0.05}$&                      &  14.11$^{+0.07}_{-0.05}$&                 &14.15$^{+0.08}_{-0.07}$\\[1.0ex]
{\OVI}~$\lambda$1032   &COS/{\it HST} &  0.233$\pm$0.012     &                         &  0.189$\pm$0.011     &                         & 0.044$\pm$0.006 &                     \\
{\OVI}~$\lambda$1038   &COS/{\it HST} &  0.105$\pm$0.010     &                         &  0.076$\pm$0.008     &                         & 0.029$\pm$0.007 &                     \\
{\bf OVI}              &              &                      &  14.49$^{+0.03}_{-0.03}$&                      &  14.33$^{+0.04}_{-0.04}$&                 &13.94$^{+0.07}_{-0.06}$\\[1.0ex]
{\bf HI}               &FOS/{\it HST} &                      &  18.3$\pm$0.3$^e$       &                      & 18.3$\pm$0.3$^e$        &                 &                       \\
                       &COS/{\it HST} &                      &                         &                      &                         &                 &16.03$^{+0.18}_{-0.18}$\\\hline 
\end{tabular}}              
\end{center}       
\begin{flushleft} 
$^a$ The equivalent width limits are computed at the 3$\sigma$-level
and are for an unresolved cloud. The column densities are measured
using the AOD method and the column density limits are computed using
the COG analysis (see text).

$^b$ $W_r$ and log$N(X)$ are quoted for absorption blueward
($W_r^{blue}$) redward ($W_r^{red}$) of the galaxy systemic
velocity. The velocity windows used to compute log$N(X)$ are $-240\leq
v_{abs} \leq -3$~{\kms} for gas blueward of the G2 systemic velocity
and $-3 \leq v_{abs} \leq 220$~{\kms} for gas redward of the G2
systemic velocity.

 $^c$ {\CII} is blended with an unknown line at $v_{abs}<-100$~{\kms},
which is why the {\CII} is expressed as a lower limit. The exact value
quoted is valid over the velocity window shown in Figure~\ref{fig:trans}.

$^d${\CIV}~$\lambda$1551 is affected by poor sky subtraction and is expressed as
a lower limit.The exact value quoted is valid over the velocity window
shown in Figure~\ref{fig:trans}.

$^e$ The {\HI} column density is adopted from \citet{churchill07} and
was determined by simultaneously fitting the {\Lya}, {\Lyb}, and the
Lyman limit absorption.
\end{flushleft}
\end{table*}                 

\subsection{Accounting for Blends in {\OVI}~$\lambda$1032 and {\CIII}~$\lambda$977}

\citet{churchill12b} performed detailed Voigt profile (VP) fits to the
Lyman series, using the {\it HST}/COS spectra, in order to determine
the gas-phase properties of the G1 absorption complex. The VP fits
were performed using our own software MINFIT \citep{cv01,cvc03}.  The
{\it HST}/COS instrumental line spread functions for the spectrograph
settings, and also for the observed wavelength of each modeled
transition, was computed by interpolating the on-line tabulated data
\citep{cos-ihb,kriss11}.

\citet{churchill12b} determined that a portion of the G1 {\Lyb}
complex was blended with the {\OVI}~$\lambda$1032 associated with G2,
while a different portion of the G1 {\Lyg} complex was blended with
{\CIII}$\lambda$977 associated with G2 as shown in
Figure~\ref{fig:blend} (the full {\Lyb} and {\Lyg} complex is not
shown since it spans roughly 1400~{\kms}).

The VP fits of the {\OVI}~$\lambda$1032 blended portion of the {\Lyb}
complex was well constrained by the non-blended {\Lya}, {\Lyg},
{\Lye}, and {\Lyf} lines. The VP fits of the {\CIII}~$\lambda$977
blended portion of the {\Lyg} complex was well constrained by the
non-blended {\Lya}, {\Lyb}, {\Lye}, and {\Lyf} lines. We show the
blended and deblended {\OVI}~$\lambda$1032 and {\CIII}~$\lambda$977
lines in Figure~\ref{fig:blend}. Given that G1s hydrogen lines are
unsaturated and can be well modeled using most of the {\HI} series, we
are confident that we have a robust correction for the spectral shape
of the {\OVI}~$\lambda$1032 and {\CIII}~$\lambda$977
lines. Furthermore, these corrections only effect regions redward of
the galaxy systemic velocity.

\subsection{Apparent Optical Depth Method}

We employed the apparent optical depth method (AOD) to measure the
column densities for each transition using the formalism of
\citet{savage91} and \citet{cv01}.  In the cases that multiple
transitions of a given ion were measured, we computed the optimal
weighted mean column density in each velocity bin.  Since the weighted
mean requires the inverse square of the uncertainties, in these cases
we treated the non-normal distributions in the uncertainties of the
optical depths using the quadratic model of
\citet{d'agostini00a,d'agostini00b,barlow03}.  The quadratic model
approximates the probability distribution of the asymmetric
uncertainties in each optical depth data point by a parabola fit to
the dimidated Gaussian constructed from the upward and downward
measured optical depth uncertainties.  Relative to a treatment that
neglects the non-normal distribution in the uncertainties, the method
effectively provides a point-by-point bias correction to the resulting
mean.

In Figure~\ref{fig:AOD} we present the optimal column density
distribution for each transition. The {\MgII}, {\CIV}, and {\OVI} AOD
distributions were computed using the doublet. For the
{\CIV}~$\lambda$1548 line, we masked out pixels that were unusable as
indicated by the two spikes in the sigma spectrum visible in
Figure~\ref{fig:trans}. Note the difference in the optical depth as a
function of velocity of each transition, especially between {\MgII}
and {\CIV}. We also employed the AOD method to compute the remaining
column densities presented in Table~\ref{tab:abs}.

Even though the Lyman series is saturated, we have optimally combined
all the series lines to produce the {\HI} column density distribution
presented in Figure~\ref{fig:AOD}. Note that some central pixels are
saturated and we therefore can not obtain a total column density,
however, we can compute the {\HI} column density over a range of
velocity windows where there is no saturation.

\begin{figure}
\includegraphics[angle=0,scale=0.44]{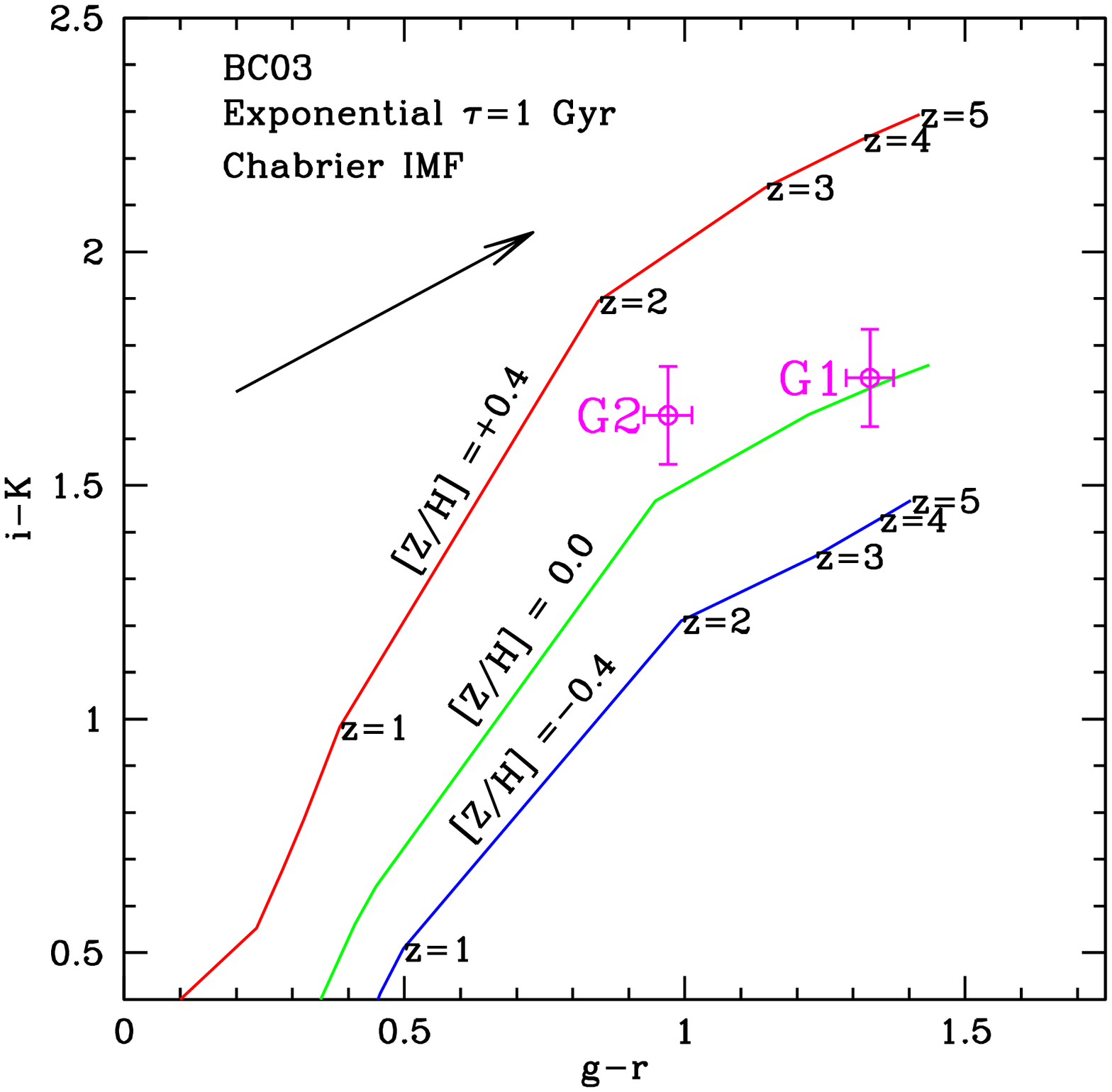}
\caption{--- Color-color diagram for $g-r$ versus $i-K$.  Lines are
  \citet{bruzual03} stellar population models for an exponential star
  formation history and a \citet{chabrier03} initial mass
  function. Stellar metallicities and formation redshifts are also
  labeled.  The points (magenta) are galaxies G1 and G2, where G2 is
  the focus of this paper and discussions of G1 can be found in a
  companion paper \citet{churchill12b}.  A reddening vector for
  $E(B-V)=0.1$ at the rest-frame of the galaxies is provided for
  reference. G2 is consistent with stellar population model of solar
  metallicity.}
\label{fig:pop}
\end{figure}

\begin{figure}
\includegraphics[angle=0,scale=0.64]{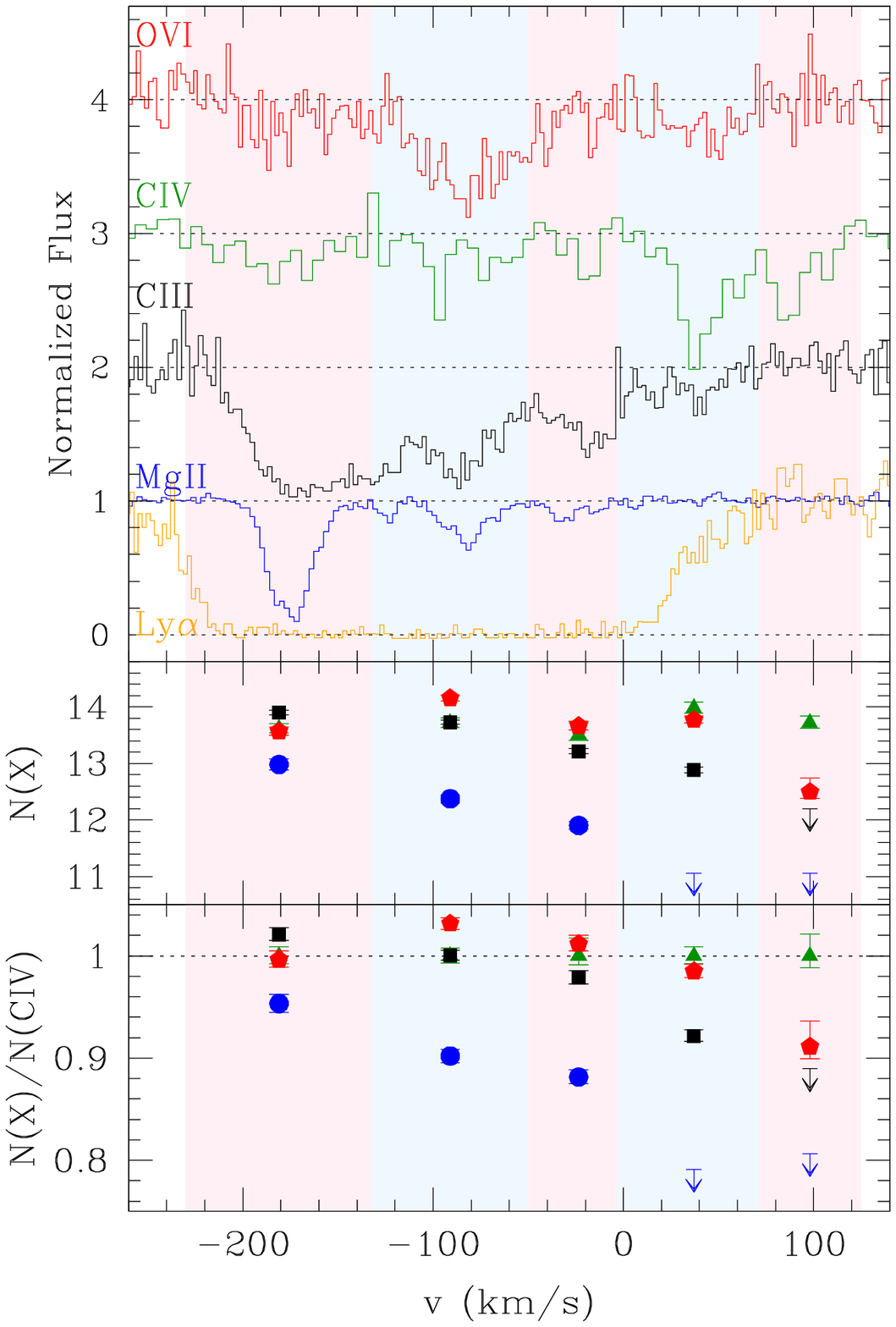}
\caption{--- (top) Overlay of the {\Lya}, {\MgII}~$\lambda$2796,
  {\CIII}~$\lambda$977, {\CIV}~$\lambda$1551 and {\OVI}~$\lambda$1032
  transitions. Note that the low ionization {\MgII}~$\lambda$2796,
  {\CIII}~$\lambda$977 trace the bulk of the hydrogen while the higher
  ionization states trace both low and high hydrogen column
  densities. The line morphologies of {\CIV} and {\OVI} are quite
  different to those of {\MgII} and {\CIII}. --- (middle) AOD column
  densities of {\MgII} (blue circles), {\CIII} (black squares), {\CIV}
  (green triangles) and {\OVI} (red pentagons) as a function of
  different velocity bins. The data points are located in the center
  of the bin, while the shading indicates the full velocity width of
  the bin. --- (bottom) Same as above except the values are normalized
  to $N$(\CIV). Note that {\CIV} and {\OVI} have similar behavior
  blueward of the systemic velocity.}
\label{fig:regions}
\end{figure}

\subsection{Curve of Growth Analysis}

We use the curve of growth analysis (COG) to determine the column
density limits of various transitions of silicon and for {\MgII},
{\MgI} and {\CII} redward of the galaxy systemic velocity.  The
equivalent width limits quoted in Table~\ref{tab:abs} are 3$\sigma$
limits for a unresolved single cloud. The single cloud assumption for
silicon is motivated by the {\SiII}~$\lambda$1527 absorption-line,
given that we would likely expect to detect the remaining silicon
transition as a single cloud component consistent with the strongest
{\MgII} cloud located at roughly $-$180~{\kms}.

The measured equivalent width limits are small and being on the linear
part of the curve of growth implies that the column densities are
mostly independent of the Doppler parameter ($b$). In fact, for $b\geq
5$, the predicted column density are independent of $b$. We find that
the {\SiII} column density is best constrained by the
{\SiII}~$\lambda$2515 data and the {\SiIV} column density is best
constrained by the {\SiIV}~$\lambda$1394 data.

There also exists {\it HST}/FOS spectra of this quasar that covers
additional silicon transitions, among other metal-lines, however, we
find that the equivalent widths were inconsistent with the predictions
by the COG analysis; this was not the case for the STIS or COS data.
We note that the FOS measurements published in
\citet{jannuzi98,archiveI} are plagued by unresolved blends from the
other {\HI} lines and therefore we do not use the FOS data in our
analysis.

It is important to state that although the curve of growth analysis
seems to systematically underestimate the column densities of the
metal-lines \citep{prochaska06}, the silicon transitions do not
provide sufficient constraints on our analysis and do not affect our
results. We provide these column density measurements for
completeness.

\section{Results: Galaxy Properties}
\label{sec:galresults}

In Figure~\ref{fig:qsofield} we present a $g'i'Ks$ color composite
centered on the quasar with galaxies G1 and G2 labeled. We find that
G2 tends to be bluer than G1, which likely indicates a younger stellar
population. This is coincident with the fact that G1 exhibits weak
(3~{\AA}) {\OII} emission.

In Figure~\ref{fig:pop}, the $g-r$ versus $i-K$ colors of G1 and G2
are compared to the \citet{bruzual03} stellar population models
obtained using EzGal \citep{mancone12} for an exponential star
formation history, with an $e$-folding time of 1~Gyr, and a
\citet{chabrier03} initial mass function.  The K-band is crucial in
breaking degeneracies between other star formation history models
since it is very sensitive to the old stellar populations. We tested a
variety of models and find that anything but a prolonged $\tau=1$~Gyr
exponential model do not fit the data.

In our companion paper \citet{churchill12b} we discuss G1, although we
note that G1 is redder, formed at an earlier epoch and is more
massive. The model comparison suggests that G2 is dominated by a
$\sim4$ Gyr stellar population with slightly greater than solar
metallicity abundance and formed at redshift $z\sim2$. We also note
while G2 has an {\OII} rest equivalent width of 3.0~{\AA}, it is
consistent with galaxies having $U-B\gtrsim 1.0$ \citep{cooper06}: red
galaxies with little-to-no star formation. Thus, the stellar
population of G2 should be dominated by an older population as the
model suggests.  Adopting the $K$-band mass-to-light ratio associated
with this stellar population model, we estimate the total stellar mass
of G2 to be $M_{\ast}\sim1\times10^{11}$~M$_{\odot}$.

We further estimate the halo virial mass from the stellar mass using
GalMass \citep{stewart11c}. GalMass uses abundance matching models
from \citet{moster10} along with semi-empirical fits to observed
galaxy gas fractions to convert between the stellar and gas mass
described in \citep{stewart09}.  There is roughly a 0.25 dex
uncertainty in $M_{\rm vir}$ at a fixed $M_{\ast}$ due to the
systematics in estimates of $M_{\ast}$ \citep{behroozi10}.  The mean
galaxy gas and baryonic masses are also estimated from
\citet{stewart11a} who employed the baryonic Tully-Fisher relation
\citep{mcgaugh05}, the stellar, gas, and dynamical mass relation
\citep{erb06}, and galaxy-gas fraction and stellar mass relation
\citep{stewart09}. We estimate the halo virial mass, gas mass and
baryonic mass to be log $M_{\rm vir}=12.9$, log $M_{\rm gas}=10.1$ log
$M_b=11.1$, respectfully.


%
%
%

\section{Results: Absorption}
\label{sec:absresults}

From Figure~\ref{fig:trans} and Figure~\ref{fig:AOD}, it is clear that
the absorption properties vary significantly as a function of
ionization level. In Figure~\ref{fig:regions}, we overlay selected
transitions to further demonstrate this. It is clear that {\MgII} and
{\CIII} (along with {\SiII} and {\MgI}) have similar kinematics and
trace the bulk of the hydrogen. While {\CIV} and {\OVI} trace some of
the same gas, their kinematics and their relative absorption strengths
differ. Furthermore, they both exist where there are diffuse hydrogen
column densities and no {\MgII} absorption.

In the bottom panels of Figure~\ref{fig:regions} we show the AOD
column densities as a function of the velocity for each transition.
Note that the {\OVI} follows similar abundances to {\CIV} across the
profile, while {\MgII} and {\CIII} differ and decrease toward the
positive velocities. It appears quite clear that the kinematics and
the abundance patterns show that {\CIV} and {\OVI} trace different
phases of gas and this difference become greater redward of the galaxy
systemic velocity where the hydrogen column density decreases by 2
dex.

Given the observed kinematic and abundance profile differences, we
have broken the absorption regions into two separate components and
model them independently. In Table~\ref{tab:abs}, we list the total
column densities for gas blueward, $N(X)^{blue}$, and redward,
$N(X)^{red}$, of the galaxy systemic velocity. In the following
subsections, we present models to explain the observed absorption
kinematics and also apply photo+collisional ionization models to
determine the origins and the physical properties of the gas.

\begin{figure*}
{\begin{tabular}{l}
\includegraphics[angle=0,scale=0.46]{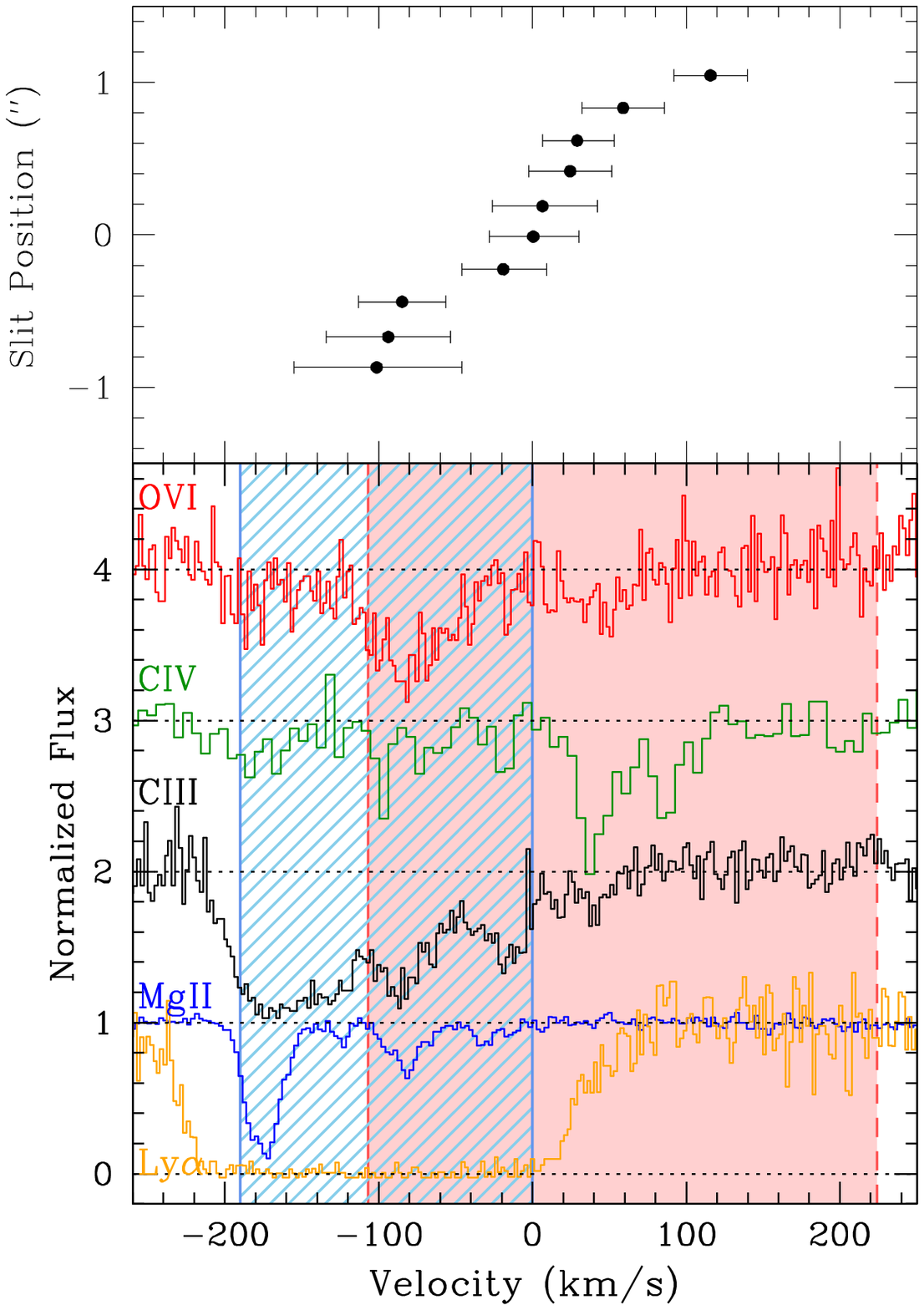}
\includegraphics[angle=0,scale=0.30]{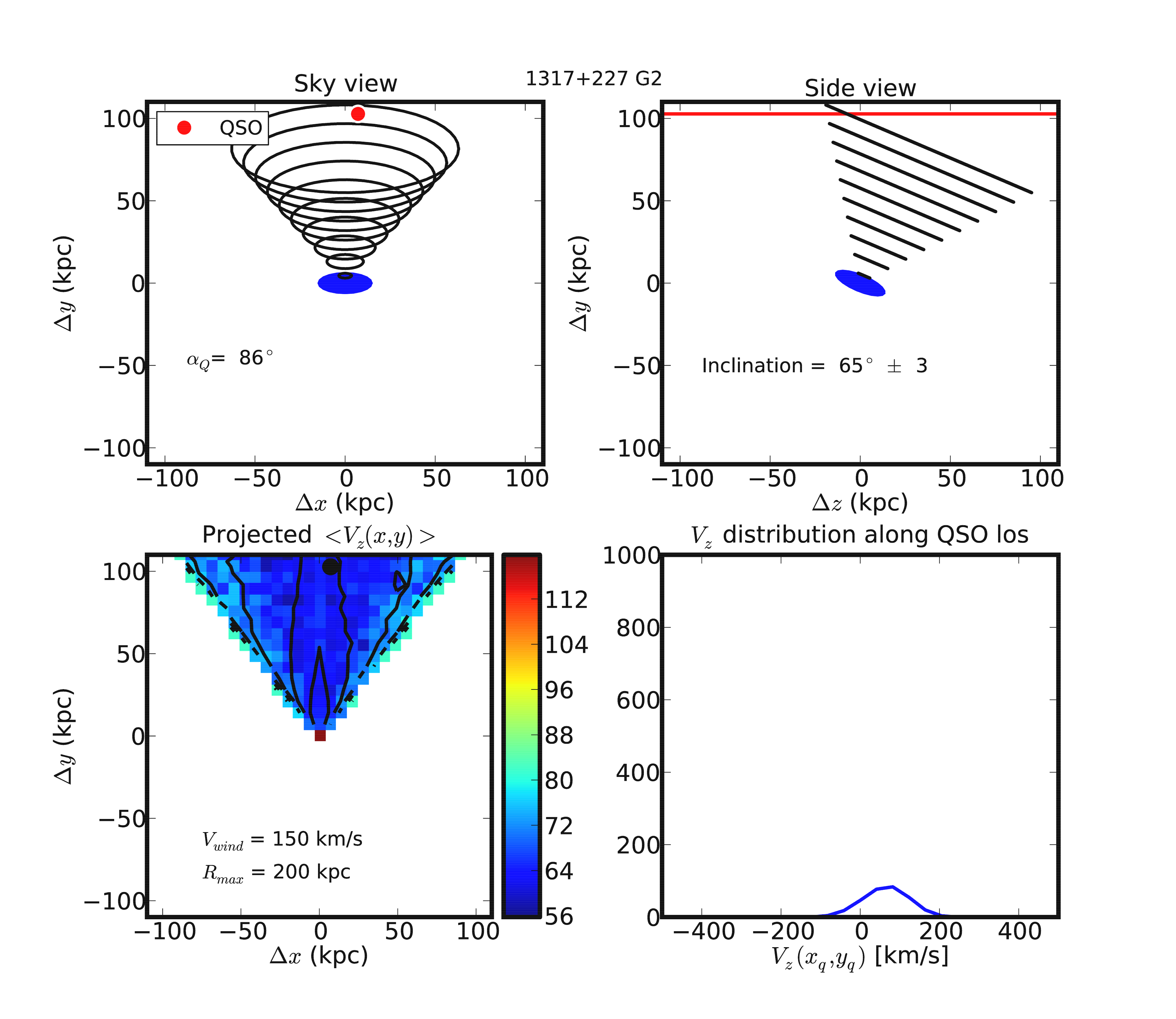}
\end{tabular}}              
\caption{--- (left) The data points show the rotation curve of G2
  obtained from \citet{steidel02}. Below are a selection of absorption
  transitions where the hashed region (blue) shows the rotating thick
  disk modeled velocities, while the solid shading (red) indicates the
  wind model velocity predictions. Note that all of the {\MgII}, and
  the majority of the {\CIII}, resides to one side of the galaxy
  systemic velocity while {\CIV}, and some {\OVI} and {\CIII} also
  resides redward of the galaxy systemic velocity. --- (right) The
  wind model of \citet{bouche11} applied to G2. --- (top-left panel)
  The wind modeled cone, having an opening angle of 30 degrees, viewed
  face-on where the galaxy is represented as the filled ellipse and
  with the QSO line-of-sight marked. --- (top-right panel) A side
  view.  --- (bottom-left panel) The average cloud line-of-sight
  velocities as a function of position within the wind. The QSO
  location is represented as the filled circle. --- (bottom-right
  panel) The line-of-sight velocity distribution of the clouds at the
  location of the quasar. We find that an outflow velocity of
  150~{\kms} produces a good match to the data and comparable to the
  results found by \citet{bouche11}. Note the peak of the density
  distribution coincides with the bulk of the {\CIV} absorption.}
\label{fig:kine}
\end{figure*}

\subsection{Kinematic Models}

\citet{steidel02} obtained the rotation curve for G2 and is
represented in Figure~\ref{fig:kine}.  Note that all of the {\MgII},
and the majority of the {\CIII}, resides to one side of the galaxy
systemic velocity. \citet{cs03} noted that the {\CIV}, and here we see
the {\OVI}, resides on both sides of the galaxy systemic velocity.

To explore the origins of the {\MgII} absorption, \citet{steidel02}
employed a simple lagging halo model that extended the galaxy rotation
velocity out into a co-rotating gaseous disk/halo.  They determined
that a lagging halo model can account all of {\MgII} absorbing gas
kinematics, even though the absorption occurs at $D=103.9$~kpc.  The
predicted model velocities are highlighted by the hashed (blue) region
below the galaxy rotation curve in Figure~\ref{fig:kine}. Further note
that the observed rotation curve does not go deep enough to reach the
flat part of the curve. This implies there is additional/higher galaxy
kinematics that were not included in the model, thus the vertical
solid (blue) line on the left of the absorption profiles would move
towards bluer velocities. It is quite clear that the low ionization
gas seems coupled to the disk kinematics similar to the models of
\citet{stewart11b}. However, the observed co-rotating {\MgII}
absorption extends to larger $D$ than the models of \citet{stewart11b}
since, at $z=1.4$, their simulated disks extend only to
$\sim$40~kpc. Thus, it remains unclear if there is size evolution as a
function of redshift or if the simulations (without metals) properly
trace the metal-lines or if this is a result of simulating only a few
galaxies.  Aside from these caveats, the similarities between the
observations and the simulations are suggestive that this absorption
system exhibits signatures of cold mode accretion.

Note that the rotating disk model does not account for the all of the
absorption. Perhaps that given that the quasar line-of-sight passes
along the minor axis of the galaxy and that G2 is undergoing some
star-formation (indicated by the {\OII} emission shown in
Figure~\ref{fig:qsofield}), then it is possible that some of the
absorption arises from outflows.

By analyzing 10 $z$$\sim$$0.1$ {\MgII} absorbers selected from
\citet{kacprzak11a}, \citet{bouche11} found a bi-modal distribution of
the azimuthal orientation of the quasar sight-lines: half of the
sight-lines aligned with the major axis and half within $\alpha=$30
degree of the minor axis. The bi-modal azimuthal angle distribution
was later confirmed by \citet{kacprzak12} using a sample of 88
absorption-selected galaxies ($W_r(2796)\geq 0.1$\AA) and 35
non-absorbing galaxies ($W_r(2796) < 0.1$\AA). These results indicate
that both gaseous disks and strong bipolar outflows could contribute
to {\MgII} cross-section. \citet{bouche11} also applied a simple
bi-conical wind model that was able to reproduce the observed {\MgII}
kinematics for the sight-lines aligned with the minor axis.  We apply
their model here in an effort to reproduce the observed absorption
kinematics. Their galactic wind model consists of $10^5$ ``clouds''
moving at a constant velocity ($V_{out}$) that is determined from the
absorption data. The clouds are contained within a cone that has a 30
degree opening angle.

Figure~\ref{fig:kine} shows the wind model for G2 using the
orientation parameters from \citet{kacprzak11b}.  The top-left panel
shows the cone view face-on and the top-right panel shows a side view
of the cone, where the galaxy is represented as the filled ellipse and
with the QSO line-of-sight marked. The bottom-left panel shows the
average cloud line-of-sight velocities a function of position. The
bottom-right panel shows the distribution of the cloud line-of-sight
velocities along the quasar sightline.  The wind speed is tuned to
match the observed velocity range. For this particular case, we find
that an outflow velocity of 150~{\kms} produces a good match to the
data and are comparable to the results found by \citet{bouche11}.

In the lower-right panel, we see the predicted absorption distribution
peaks at roughly 50~{\kms}, where the majority of the {\CIV} and some
{\OVI} resides. The wind model velocity range is also shown as the
solid shaded region over the absorption profiles. The model could also
account for the observed {\CIV} and {\OVI} blueward of the galaxy
systemic velocity, although the model predicts only predicts of few
percent of the gas is expected at these velocities.

If the opening angle or the wind speed is increased, the model
velocity range would also increase and could include all of the
observed absorption. However, increasing these parameters will not
reproduce the observed optical depth distribution of the cold and hot
gas. While our current model is tuned to reproduce the absorption
reward of the galaxy systemic, a different model would not be able to
predict the strong absorption residing at -200~{\kms}. A model with
increased opening angle and/or wind speed would also predict
additional absorption beyond $+200$~{\kms} where none is
observed. Therefore, it is very difficult to reproduce both the
velocity and optical depth distributions of the cold and hot gas using
only the wind model.

The wind model can not account for the bulk of the {\MgII}, {\MgI},
{\SiII} and {\CIII}. Thus, although the galaxy is forming stars, winds
are likely not responsible for the cool gas, but could be responsible
for the hot gas. One would naively expect that the infalling gas would
be metal poor while the outflowing gas to be metal enriched. We
explore the gas-phase metallicities in the next subsections in order
to determine the possible origins of the absorption.

\begin{figure}
{\begin{tabular}{l}
\includegraphics[angle=0,scale=0.41]{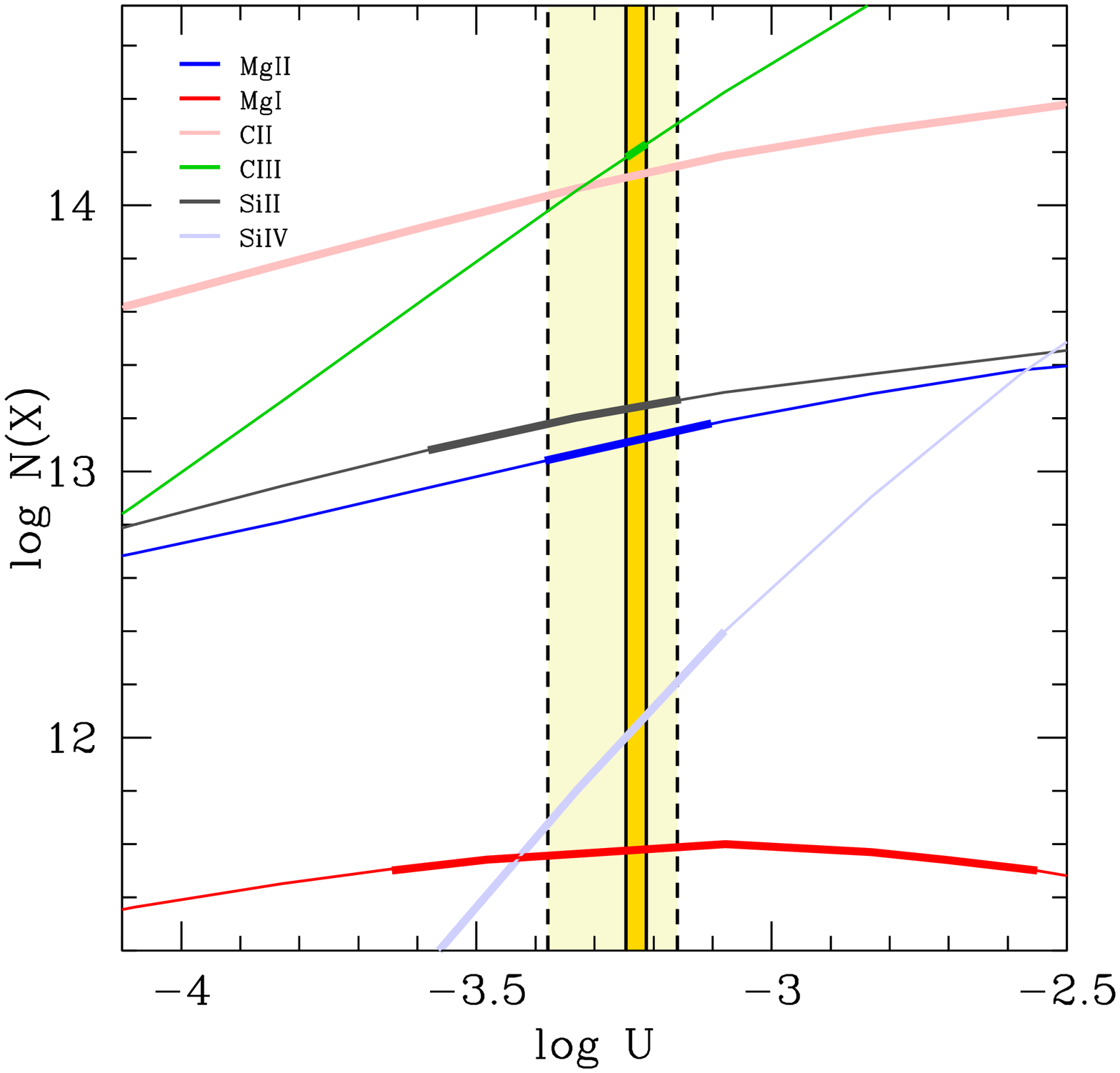}\\
\includegraphics[angle=0,scale=0.41]{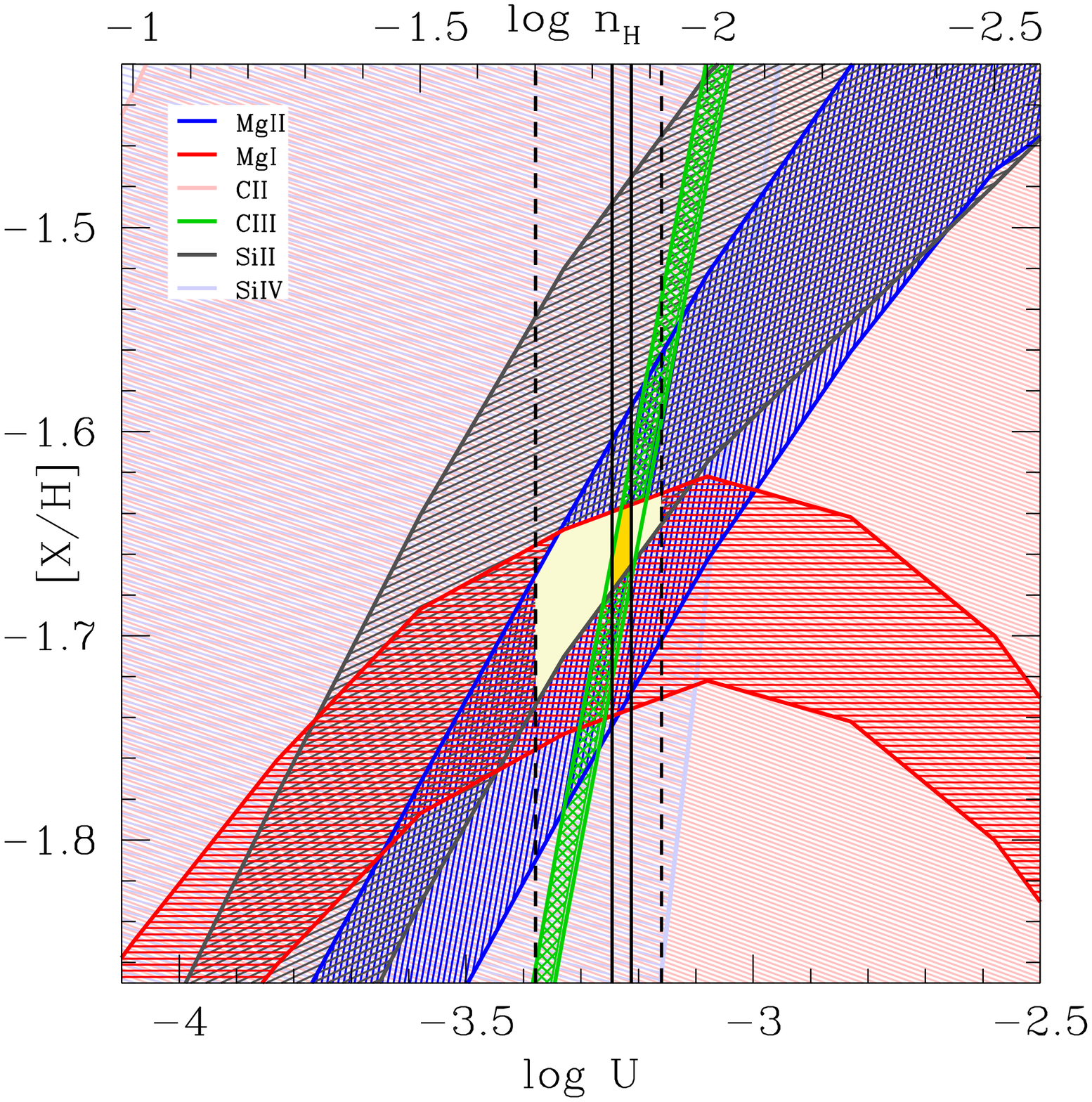} 
\end{tabular}}              
\caption{--- (top) The Cloudy predicted column densities, as a
function of the ionization parameter, for a metallicity of
$[X/H]=-1.67$ and a log[$N$(\HI)]=18.3. The bold portions of the
curves show the observed column densities, $N(X)^{blue}$, blueward of
the galaxy systemic velocity (Table~\ref{tab:abs} column 6). The
vertical lines show the ionization parameter constraints provided by
the data.  The {\CIII} provides the tightest constraints of $-3.25\leq
{\rm log}U \leq -3.21$ indicated between the solid vertical lines
(gold region). {\SiII} and {\MgII} provide secondary constraints of
$-3.38\leq {\rm log}U \leq -3.16$ as noted by the vertical dashed
lines (yellow region). --- (bottom) The ionization parameter as a
function of metallicity for the above model.  The shaded regions show
the model ranges due to the measurement errors in $N(X)^{blue}$. The
vertical lines are the ionization parameter constraints provided by
the top panel and provides the region of acceptable metallicities. The
overlapping shaded regions of each ion within the constrains provided
by $U$ have a solid shading and outlines the allowed metallicities of
this cold gas. Again, {\CIII} places the tightest constraints on the
metallicity $-1.68\leq [X/H] \leq-1.64$.  Thus, the cool gas has low
metallicity and a low ionization parameter.}
\label{fig:diskmodel}
\end{figure}

\begin{figure}
\includegraphics[angle=0,scale=0.80]{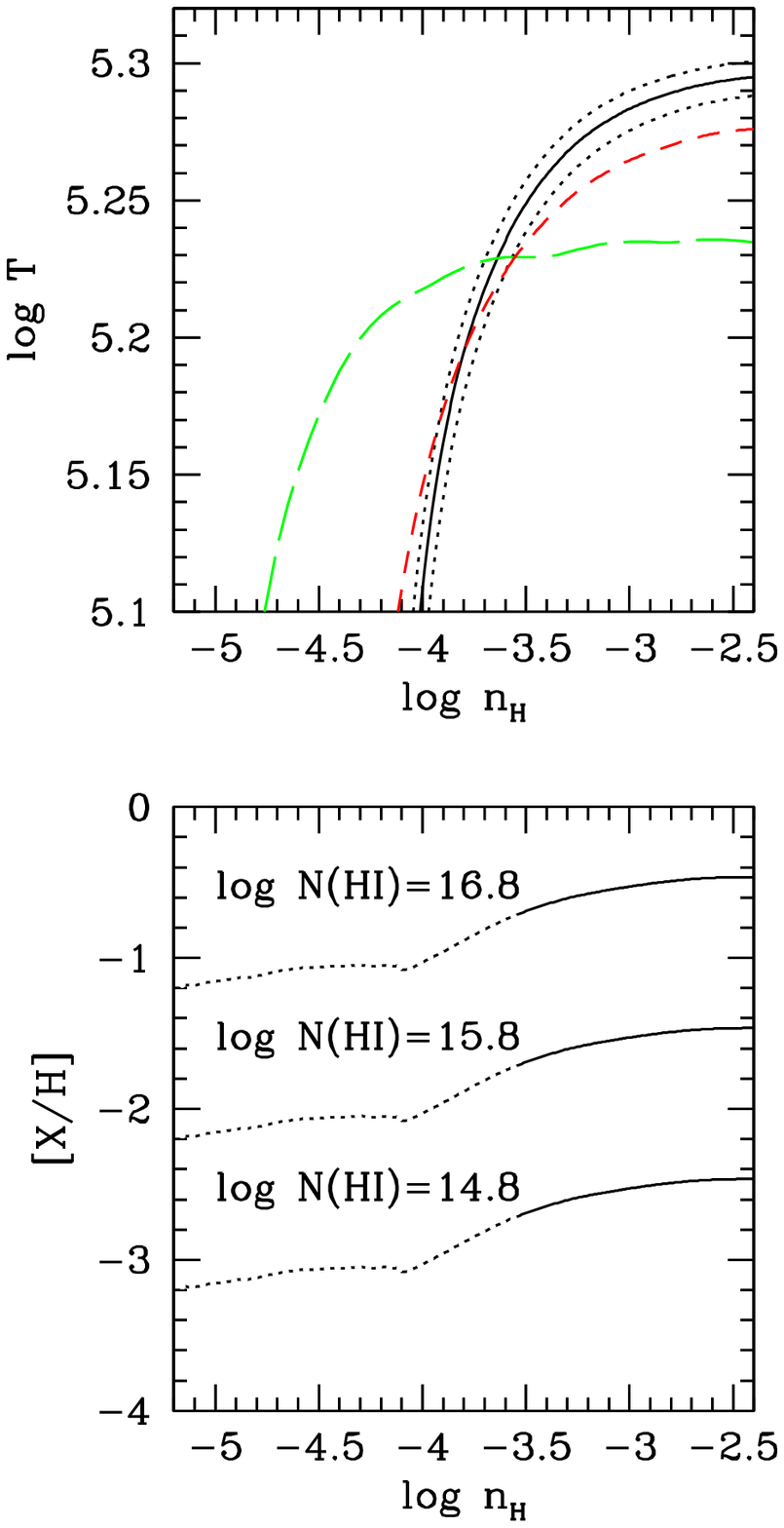}
\caption{--- (top) The photo+collisional ionization models
\citep{churchill12c} as a function of temperature and hydrogen
density. The lines indicate abundance ratios which are independent of
the hydrogen column density. The solid line is the {\CIV}/{\OVI}
column density ratio along with its measured error (dotted lines). The
long and short dashed lines are the {\CIII}/{\CIV} and {\CIII}/{\OVI}
column density ratio, respectively. The {\CIII}/{\CIV} and
{\CIII}/{\OVI} dashed curves are lower limits on the allowed
temperature range for a given hydrogen density.  Note that full self
consistency is when the dashed curves to fall below the black
curve. The gas has a hydrogen density of log$n_{_{\rm H}}>-3.5$ and the
{\CIV}/{\OVI} provides a temperature constraint of log$T =
5.23-5.29$. --- (bottom) The metallicity as a function of hydrogen
density for various hydrogen column densities. We provide conservative
limit of log$N(\HI)<16.8$ [see text] which yields $[X/H]<-0.5$.  }
\label{fig:wb}
\end{figure}

\begin{figure}
\includegraphics[angle=0,scale=0.91]{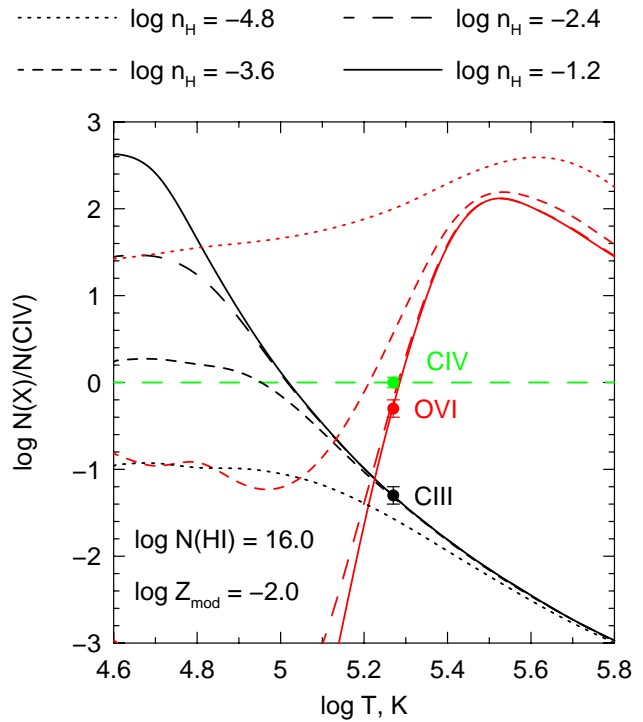}
\caption{--- The photo+collisional ionization models obtained from
\citet{churchill12c} as a function of temperature and column density
normalized to $N(\CIV)$. The data points are the AOD measured column
densities of absorption redward of G2's systemic velocity,
$N(x)^{red}$, shown in Table~\ref{tab:abs} (column 8).  We allowed the
gas temperature and metallicity to vary for a range of hydrogen
densities, $n_{_{\rm H}}$.  The data constrain the gas temperature to
be 185,000K and limits log$n_{_{\rm H}}\geq-2.4$. In
Figure~\ref{fig:Zconst} we provide the constraints on the warm phase
metallicity.}
\label{fig:Tconst}
\end{figure}

\subsection{Cold Gas Phase}

To determine the physical properties of the cool gas blueward of the
galaxy systemic velocity, we use Cloudy \citep{ferland98} to model the
ionization conditions. We follow the standard assumption of a
photoionized uniform slab of gas that is in ionization equilibrium and
is illuminated with a \citet{haardt11} ionizing spectrum, where the UV
photons arise from quasars and galaxies.  The ionization parameters,
$U$, and the metallicity of the gas are varied to match the
observations of $N(X)^{blue}$ in Table~\ref{tab:abs} (column 6).

In Figure~\ref{fig:diskmodel} we show the model results computed for
$\log[N(${\HI}$)]=18.3\pm0.3$ and a metallicity $[X/H]=-1.67$. We find
a narrow range of ionization parameters that reproduce the cool gas
phase. The thin curves show the models while the thicker curves show
the model values permitted by the data. The {\CIII} provides the
tightest constraints on the ionization parameter of $-3.25\leq {\rm
log} U \leq -3.21$. Less stringent constraints are also found from
{\SiII} and {\MgII} yields $-3.38\leq {\rm log} U \leq -3.16$.

In Figure~\ref{fig:diskmodel}, the same Cloudy models are shown as a
function of $U$ and metallicity. The constraints placed on $U$ and the
models confine the allowed range of metallicity. The {\CIII}
measurements limit the metallicity to $-1.68 \leq [X/H] \leq
-1.64$. Recall that the {\CIII} column density measurement blueward of
the galaxy systemic velocity is the unblended portion of this
transition. If we were to include the small contribution of the
{\CIII} redward of the galaxy systemic velocity our results would
still be consistent. The {\SiII} and {\MgII} yields a less stringent
of $-1.73 \leq [X/H] \leq-1.63$. Thus, this cool gas component has low
metallicity and a low ionization parameter.

In summary, the {\CIII} column density provides the tightest
constraint and yields a log$U$=$-3.23\pm0.2$ and $[X/H]=-1.66\pm0.02$.
The gas is primarily ionized since the ionization fraction
$X$(\HI)=$N$(\HI)/$N$({\rm H})=$0.088\pm0.002$ with a hydrogen number
density of log($n_{_{\rm H}}$)=$-1.85\pm0.02$. The physical size of
the cloud is $L=N_{_{\rm H}}/n_{_{\rm H}}=$$521\pm2$~pc.

Note that {\CIV} and {\OVI} do not appear on these plots. The column
densities predict for this photo-ionized modeled gas would be
log$N({\CIV})=11.88-11.94$ and log$N({\OVI})=7.29-7.35$. This
predicted column densities are no where near the measured values of
log$N$(\CIV)=14.11 and log$N$({\OVI})=14.33. Thus, it is likely that
these ions are not part of the cool gas phase and are likely part of a
separate ``warm/hot'' collisional ionized gas phase. In
Figure~\ref{fig:regions}, note that the {\CIV} and {\OVI} have
different kinematics that the lower ionization species further
indicated that the absorption blueward of the galaxy systemic velocity
is probing two gas phases.  In the next section we model the warm gas.



\subsection{Warm Gas Phase}
\begin{figure*}
\includegraphics[angle=0,scale=0.74]{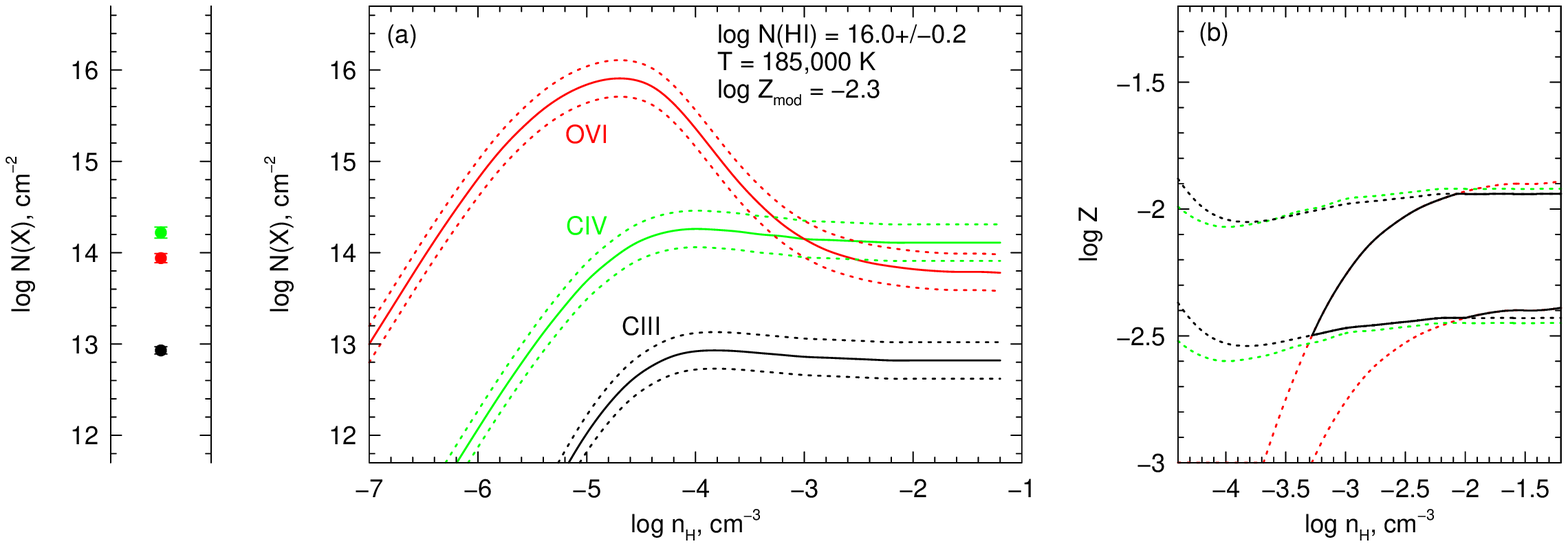}
\caption{--- (left) The data points are the AOD measured column
densities of absorption redward of G2's systemic velocity,
$N(x)^{red}$, shown in Table~\ref{tab:abs} (column 8). --- (middle)
The photo+collisional ionization models obtained from
\citet{churchill12c} for {\CIII}, {\CIV} and {\OVI} using a fixed
temperature of 185,000~K and $[X/H]=-2.3$. The spread in the models is
due to the error in the measured {\HI} column density of
log[N(\HI)]=16.03$\pm 0.18$ (dotted lines). --- (right) The allowed
metallicity and $ n_{_{\rm H}}$ for the model on the left. While
$n_{_{\rm H}}$ is not well constrained, we find $-2.50\leq [X/H] \leq
-1.93$ and log$n_{_{\rm H}} > -3.3$ for the lower metallicity limit
and log$n_{_{\rm H}} > -2.1$ for the upper metallicity limit.}
\label{fig:Zconst}
\end{figure*}

We employed our own photo+collisional ionization code
\citep{churchill12c} to model the warm gas phase since it is optimally
designed for optically thin gas with no ionization structure. In
short, the code incorporates photoionization, Auger ionization, direct
collisional ionization, excitation-autoionization,
photo-recombination, high/low temperature dielectronic recombination,
charge transfer ionization by H$^{+}$, and charge transfer
recombination by H$^0$ and He$^0$. All metal transitions and
ionization stages for elements up to zinc are modeled. Solar abundance
mass fractions are obtained from \citet{draine11} and
\citet{asplund09} and a \citet{haardt11} ionizing spectrum is used for
the ultraviolet background. The model inputs are the hydrogen number
density ($n_{_{\rm H}}$), kinetic temperature and the gas metallicity
while the model outputs are the electron density, and the ionization
and recombination rate coefficients, ionization fractions and the
number densities for all ionic species.  Our models are consistent
with those of Cloudy for log($n_{_{\rm H}}$)$>-3.75$. For
log($n_{_{\rm H}}$)$<-3.75$, cosmic ray heating in Cloudy tends to
yield over-ionized clouds with large sizes ($<100$~kpc). When cosmic
ray heating is turned off, Cloudy has convergence issues and tends to
produce ratios of column densities for adjacent species that do not
follow a trend physical with ionization potential.  We have reconciled
these issues in our models and a detailed comparison will be presented
in \citet{churchill12c}.

\subsubsection{Warm Gas Phase Blueward of Systemic Velocity}
\label{sec:warmblue}

In modeling the cold gas phase we noted that we are unable to account
for the measured logN(\CIV)=14.11 and the logN({\OVI})=14.33. Thus,
these ions are not part of the cool gas phase and are likely part of a
separate warm collisionally ionized gas phase. Here we model the
diffuse warm gas found blueward of G2's systemic velocity.
Furthermore, all of the measured {\CIII} is well modeled as being
associated with the cool gas phase, however, there could possibly
exist a small fraction of a measurable warm {\CIII} component
hidden. If we assume that the warm component is within the 0.2 dex
measurement error of {\CIII}, then there can be at most
log$N$({\CIII})$<$12.87 in the warm phase.

In Figure~\ref{fig:wb} we show column density ratios, which is
independent of $N({\HI})$, as a function of temperature and hydrogen
density. The solid line is the {\CIV}/{\OVI} column density ratio
along with its measured error (dotted lines). The long and short
dashed lines are the {\CIII}/{\CIV} and {\CIII}/{\OVI} column density
ratios, respectfully. The dashed curves are lower limits on the
allowed temperature range for a given hydrogen density since
log$N$({\CIII})$<$12.87. Note that full self consistency is when the
dashed curves fall below the solid curve. The models indicated that
the gas has a hydrogen density of log$n_{_{\rm H}}>-3.5$ and the
{\CIV}/{\OVI} provides temperature constraints of log$T = 5.23-5.29$,
where the exact temperature is dependent on $n_{_{\rm H}}$.

In Figure~\ref{fig:wb}, we also show the gas metallicity as a function
of $N({\HI})$ and $n_{_{\rm H}}$. We are unable to measure the {\HI}
column density associated with this warm component. However,
\citet{churchill07} states that if an additional hydrogen component
having log$N({\HI})>$16.8 was added to the total hydrogen column
density, then it would measurably modify the shape of the Lyman limit.
Thus, this provides a conservative estimate of the additional
$N({\HI})$ that could be associated with the warm component since we
assume it can not contribute to the Lyman limit.  The column density
limits provides a metallicity limit of $[X/H] \lesssim -$0.5. Note
that the metallicity scales with hydrogen column density by the same
amount. It is likely that the metallicity is much lower, though we can
not provide any further constraints.

\subsubsection{Warm Gas Phase Redward of Systemic Velocity}

Here we model the diffuse warm gas found redward of G2's systemic
velocity.  We allowed the gas temperature, the gas metallicity and the
hydrogen density to vary to match the column density measurements,
$N(X)^{red}$, from Table~\ref{tab:abs} (column 8). We only detect
{\CIII}, {\CIV} and {\OVI} redward of the galaxy systemic
velocity. The {\HI} associated with the warm gas phase was computed
from the AOD {\HI} column density redward of the galaxy systemic
velocity. As shown in Figure~\ref{fig:AOD}, the {\HI} absorption
profile is not saturated redward of the galaxy systemic velocity and
contains a logN(\HI)=16.03$\pm 0.18$.

Figure~\ref{fig:Tconst} shows the models column densities normalized
to that of {\CIV}. The data are well constrained with the warm phase
gas having a temperature of $T=$185,000~K while there seems to be a
larger range in the hydrogen density such that log$n_{_{\rm
H}}\geq-2.4$ for a fixed metallicity.  In order to accurately
determine the range of metallicities and $n_{_{\rm H}}$, we show the
predicted column densities and metallicities as a function of
$n_{_{\rm H}}$ in Figure~\ref{fig:Zconst}. Again, the measured (left)
and modeled (middle) column densities are shown. The model spread,
indicated by the dotted lines, is dominated by the error in the
measured {\HI} column density. We find that the models do not well
constrain the $n_{_{\rm H}}$ but provide reasonable constraints on the
metallicity. We find $-2.50\leq [X/H] \leq -1.93$ and log$n_{_{\rm H}}
> -3.3$ for the lower metallicity limit and log$n_{_{\rm H}} > -2.1$
for the upper metallicity limit.


\begin{table*}
\begin{center}
  \caption{Modeled Absorbing Gas Properties}
  \vspace{-0.5em}
\label{tab:results}
{\footnotesize\begin{tabular}{lcccccc}\hline
Name     & velocity [{\kms}]                     &  log$N(\HI)$ &  log($T$)    &   log($U$)                                     & log($n_{_{\rm H}}$) &  $[X/H]$ \\\hline
cold     & $-240\leq v_{abs} \leq -3$            & 18.3$\pm$0.3\phantom{1}     &$3.82-5.23$& $-3.25\leq$log($U$) $\leq -3.21$   & $-1.87\leq$ log($n_{_{\rm H}}$)$ \leq -1.83$ &$-1.68\leq [X/H] \leq -1.64$\\
warm blue& $-240\leq v_{abs} \leq -3$            & $<$16.8$^a$\phantom{000001} &$5.23-5.29$& \phantom{000000}log($U$) $\geq-3.5$& \phantom{0000001}log($n_{_{\rm H}}$) $ >-3.5 $    &  \phantom{000000.} $[X/H]\leq-0.5$$^a$\\
warm red & $\phantom{24}-3\leq v_{abs} \leq 220$ & 16.0$\pm$0.18& 5.27         & \phantom{000000}log($U$) $\geq-3.3$            & \phantom{0000001}log($n_{_{\rm H}}$) $ >-3.3 $   &   $-2.50\leq [X/H] \leq -1.93$       \\\hline 
\end{tabular}}              
\end{center}       
\begin{flushleft} 
$^a$ See text in Section~\ref{sec:warmblue} for a  discussion of the
$N(\HI)$ upper limit used and how that affects the absorption metallicity.
\end{flushleft}
\end{table*}                 

\section{Discussion}
\label{sec:dis}

The galaxy G2 has a stellar metallicity greater than solar, yet it
contains $[X/H]\sim -2$ halo gas detected in absorption at 104~kpc
from the galaxy center. The data likely suggests that the absorbing
gas is tracing cold accretion. A summary of the absorption gas
properties is presented in Table~\ref{tab:results}. The kinematic
differences\footnote{\citet{ding05} modeled the {\CIV} and lower
ionization states of {\MgII}, {\MgI} and {\FeII} with the same
kinematic structure. It is clear from the fit residuals that the
kinematics are not well modeled with similar velocity structure
although they do have some similarities.}  seen between the high and
low ionization states, as shown in Figure~\ref{fig:regions}, and the
photo+collisional ionization models suggest that the absorption is
multi-phase. This may imply that the warm and cold gas physically
arises in different locations or that the absorbing gas is not well
mixed.
 
The quasar sight-line probes absorption within three degrees of the
projected minor axis of G2 \citep{kacprzak11b}, which is a location
recently interpreted to be favorable for probing galactic-scale winds
\citep{bouche11, bordoloi11,kacprzak12}. G2 is dominated by a
relatively old stellar population, yet it has measurable {\OII}
emission (3.0~{\AA}) and thus is likely undergoing some current
star-formation that could possibly produce winds. 

In Figure~\ref{fig:kine} we show a wind model that can account for the
bulk of the warm gas kinematics, both redward and blueward of G2's
systemic velocity, yet it does not reproduce the kinematics of the
cold gas. This might imply that some of the warm gas could originate
from winds, however, the $2-2.5$ order of magnitude difference between
the galaxy stellar metallicity and the absorption metallicity is
inconsistent with a wind model \citep[e.g.,][]{oppenheimer10}.
Assuming a constant wind speed of 150~{\kms}, as suggested by the
models, the absorbing gas would have left the galaxy 670~Myr
ago. Given such a short time scale, G2 still expected to have solar
metallicity at that epoch according to mass-gas metallicity relations
\citep[e.g.,][]{savaglio05} and simulations
\citep[e.g.,][]{oppenheimer10}. The recycling for wind material is
less than 1 Gyr \citep{oppenheimer10}, thus, if the gas were wind
material and was well mixed, it would be expected to have a
metallicity similar to that of the parent stellar population. However,
the efficiency at which the gas is mixed within the halo environment
is still unclear.  \citep{sijacki11} shows that poor gas mixing in
simulations may be an artifact the suppression of dynamical fluid
instabilities demonstrated by comparing smooth particle hydrodynamics
codes to that of the moving mesh code AREPO. However, some LLSs
exhibit metallicity variations across the absorption profile
\citep[e.g.,][]{prochter10}.  Thus, if the gas probed by this
absorption system is poorly mixed, it is plausible that the gas is
entrained in a outflowing wind \citep[see][]{auzas12,schaye07}.
Although, the current observational evidence suggests that {\MgII}
entrained in outflowing material may, on average, have a maximum
projected extension of $\sim 50$ kpc.  \citep{bordoloi11}. Thus, the
combination of the low metallicity, the relative kinematics between
the galaxy and the absorbing gas, and the large projected distance
between the gas and the galaxy, corroborate an accretion scenario.

In summary, although the kinematic model works well for the warm gas,
it is inconsistent with the expected outflow metallicities and this
gas would have already been recycled several times and should have
near solar metallicity. This leads to the idea that this meal poor gas
could be infalling towards the galaxy.


Galaxy G2 has a stellar mass of $M_{\ast}=1\times10^{11}$~M$_{\odot}$
and a viral mass of $M_{\rm vir}=8\times10^{12}$~M$_{\odot}$.  Cold
mode accretion is less favored for galaxies with $M_{\rm vir} \gtrsim
10^{12}$~M$_{\odot}$ since the cold gas can become shock-heated as it
enters the halo and hot mode accretion becomes the dominant mode of
accretion
\citep[e.g,][]{dekel06,keres09,stewart11a,stewart11b,freeke11a,freeke11b}.
For massive galaxies, infalling gas can be shock heated to
temperatures above $10^6$~K, however, accretion via dense filaments
can be maintained past the shock regions, although have dramatically
decreased covering fractions compared to low mass galaxies, and still
accret onto the galaxy
\citep[e.g,][]{keres09,stewart11a,stewart11b,faucher-giguere11,freeke11a,freeke11b}.
The hot gas pressure compresses cold streams and provides an efficient
means of bringing cold pristine gas to the host galaxy and should be
traced by LLS \citep[e.g.,][]{fumagalli11}.  A $\log N({\HI}) = 18.3$
corresponds to $\Delta \rho/\rho > 1000$ at $z \leq 1$
\citep[e.g.,][]{dave99} indicates that these high densities could make
its way into the galaxy center.

Although massive galaxy halos are built up from hot accretion,
comprising of $80-90$\% of the total accretion, it becomes less
important for accretion on to the ISM [$50-70$\% hot mode]
\citep{freeke11a}.  However, \citet{freeke11a} states that the cold
gas accreting onto the ISM in massive galaxies increases with
increasing simulation resolution. This could arise since gas clouds
could be easily disrupted in SPH simulations, and the cold fraction
would increase if higher densities were reached in higher resolution
simulations.  A lower hot fraction of gas accreted onto the galaxy
occurs because the hot gas temperature increases with the viral
temperature, or halo mass, resulting in longer cooling times
\citep[e.g.,][]{wiersma09} and yielding less hot gas cooling to ISM
temperatures.  However these results could also be dependent on the
type of simulations used \citep{sijacki11}.

Using the cosmological formulae for virial quantities of
\citet{bryan98}, we derived the radius, $R_{\rm vir}= 380$~kpc, the
circular velocity $v_{\rm circ}=280$~{\kms} and the temperature,
$T_{\rm vir}= 3\times 10^6$~K. The quasar is probing along the minor
axis of the galaxy within the viral radius at $R/R_{\rm vir}\sim0.3$,
well within the shock radius. Furthermore, the halo gas is expected to
be heated to the viral temperature. The temperatures deduced from our
models for the warm gas phase is $1.8\times 10^5$~K for gas blueward
of the systemic velocity and $1.9\times 10^5$~K for the gas redward of
the galaxy systemic velocity: both are an order of magnitude cooler
than that of the viral temperature. However, this gas can possibly
cool after it had been shocked, although as mentioned above, the
cooling time for hot shock-heated gas is much longer for massive
galaxies.

We can estimate a cool gas phase temperature by using the Doppler
parameters derived from {\MgII} absorption \citep{kacprzak11b}. The
{\MgII} Doppler parameters range from $2.1-10.8$~{\kms} and if we
assume that all the broadening is thermal, we can compute upper limits
on the temperature of $6.6\times 10^3-1.7\times 10^5$~K.  These
temperatures are upper limits since some of the line-broadening could
be due to turbulence, and the Voigt profiles fits to the data assume a
minimum number of ``clouds''; if more clouds were inserted, than the
velocity width of each line would decrease.  Thus, the cold gas has
temperatures well expected for cold mode accretion and not post-shock
heated cooling gas accreting within a viral radius ratio of 0.3
\citep{freeke11b}.

The absorption kinematics also hints to an origin of cold-mode
accretion. The metal poor cold-phase, and redward warm-phase, has
kinematics consistent with extended disk rotation \citep{kacprzak10a,
  steidel02}. This result has been interpreted using cosmological
simulations to be a signature of cold gas accreting via filaments that
drive the angular moment of the galaxy, thereby mimicking its rotation
out at larger impact parameters \citep{kacprzak10a, stewart11b}. This
is consistent with \citet{freeke11b} that shows cold accretion gas
having a higher radial velocities and scales with increases mass
compared to the flat radial velocity distribution of the hot-mode
accretion as a function of mass.

As we previously mentioned, the metallicities of the warm ($[X/H]=-2$
to $-2.5$) and cold ($[X/H]=-1.7$) gas are low and appear to be
consistent with metallicities expected for cold-mode accretion at
$R/R_{\rm vir}\sim0.3$ \citep{freeke11b} and do not mimic the
metallicities expected for the ISM, the hot halo, and the host galaxy
\citep{oppenheimer10}. The temperatures and kinematics of the
absorbing gas is also consistent with what is expected for cold-mode
accretion.  Thus, it is likely that the absorption is probing cold
pristine gas infalling towards the center of the disk, further fueling
star formation.


\section{Conclusions}
\label{sec:conclusion}

In this paper, we present detailed photo+collisional ionization models
and kinematics models of the multi-phase absorbing gas, detected
within the {\it HST}/COS, {\it HST}/STIS, and Keck/HIRES spectra of
the background quasar TON 153, associated with star-forming spiral
galaxy at $z=0.6610$. The sightline probes the projected minor axis of
the galaxy at projected distance of 0.3 virial radii, well inside the
virial shock radius predicted for a galaxy of this mass, implying that
if the gas is infalling that it is post shock heated accretion or a
cold filament. We obtained followup {\it HST}/COS data to study other
metal-lines in order to determine the halo gas properties and their
origins. This galaxy was targeted as a candidate cold accretion probe
supported by kinematic and orientation results presented by
\cite{steidel02, kacprzak10a, kacprzak11b}.

Our main results can be summarized as follows:

\begin{enumerate}

\item From $g'r'i'Ks$ photometry and stellar population models, we
 determined that G2 is dominated by a $\sim4$ Gyr stellar population
 with slightly greater than solar metallicity abundance and formed at
 redshift $z\sim2$. We estimate an
 $M_{\ast}=1\times10^{11}$~M$_{\odot}$ implying an log $M_{\rm
 vir}=12.9$.

\item The low ionization states, {\MgI}, {\SiII}, {\MgII} and {\CIII},
have similar absorption kinematics, abundance ratios across the
profile, and trace the bulk of the hydrogen, while {\CIV} and {\OVI}
trace some of the same gas, their kinematics, abundance ratios, and
their relative absorption strengths differ.  We infer that the low and
high ionization states trace different gas phases.

\item Modeling the cold gas blueward of G2's systemic velocity,
$N(X)^{blue}$, we constrain log$T=3.82-5.23$, $-3.25\leq$log($U$)
$\leq -3.21$, $-1.87\leq$ log($n_{_{\rm H}}$)$ \leq -1.83$, and
$-1.68\leq [X/H] \leq -1.64$. The gas is cold and very metal poor:
consistent with cold accretion.  We are unable to account for the
measured N(\CIV) and N({\OVI}) when modeling the cold phase, thus,
these ions are likely part of a separate warm collisionally ionized
gas phase.

\item A lagging halo model can account for all of low ionization
 absorption, hinting that this gas is coupled to the disk and
 simulations interpret this as a detection cold mode accretion.

\item Modeling the warm gas blueward of G2's systemic velocity,
$N(X)^{blue}$, we find $n_{_{\rm H}}>-3.5$ and log$T =
5.23-5.29$. Armed with only a conservative limit of hydrogen column
density that could be associated with the warm component
[$N({\HI})>$16.8], we estimate $[X/H] \lesssim -$0.5, although it is
highly likely that the metallicity is much lower.

\item Modeling the warm gas redward of G2's systemic velocity,
 $N(X)^{red}$, we find hot and metal poor gas with $T=$185,000~K,
 $-2.50\leq [X/H] \leq -1.93$ and $n_{_{\rm H}} > -3.3$.

\item The quasar line-of-sight passes along G2's minor axis and a wind
model can account for the observed {\CIV} and {\OVI} redward and
blueward of the galaxy systemic velocity. However, given the $2-2.5$
order of magnitude difference between the galaxy stellar metallicity
and the absorption metallicity demonstrates the gas can not arise from
galactic winds.
\end{enumerate}

It remains plausible that this low metallicity gas arises from
unidentified satellites around the host galaxy or from the incomplete
mixing between metal enriched and metal poor halo gas, However, the
combination of the relative kinematics, temperatures, and relative
metallicities allows us to conclude that the multi-phase gas detected
in absorption likely arises from cold accretion around this massive
galaxy.  For high mass galaxies the cold accretion cross-section is
expected to be a few percent, so our absorption system and others
cited in the literature could be a by-chance low probability
intersection of a filament, or the resolution effects in the
simulations \citep[see][]{freeke11b} are underestimating the covering
fraction of cold flows.  This system also contradicts current results
that predict that all absorption detected in quasars probing gas along
the projected minor axis of galaxies is produced by winds
\citep{bordoloi11, bouche11,kacprzak12}: This is clearly not the case
here.

\section*{Acknowledgments}

We thank Nicolas Bouch\'{e} for his useful comments, models and for
carefully reading this paper. We also thank the anonymous referee for
carefully reading the manuscript and for providing insightful
comments. CWC was partially support through grant HST-GO-11667.01-A
provided by NASA via the Space Telescope Science Institute, which is
operated by the Association of Universities for Research in Astronomy,
Inc., under NASA contract NAS 5-26555.  CWC thanks GGK, and Michael
T. Murphy, and Swinburne Faculty Research Grants for providing funding
for a visit to Swinburne University of Technology. Based on
observations made with the NASA/ESA {\it Hubble Space Telescope} (PID
11667), and obtained from the Hubble Legacy Archive, which is a
collaboration between the Space Telescope Science Institute
(STScI/NASA), the Space Telescope European Coordinating Facility
(ST-ECF/ESA) and the Canadian Astronomy Data Centre (CADC/NRC/CSA).
Some of the data presented herein were obtained at the W.M. Keck
Observatory, which is operated as a scientific partnership among the
California Institute of Technology, the University of California and
the National Aeronautics and Space Administration. The Observatory was
made possible by the generous financial support of the W.M. Keck
Foundation. This work is also based on observations obtained with the
Apache Point Observatory 3.5-meter telescope, which is owned and
operated by the Astrophysical Research Consortium.  Observations were
also made with the NASA/ESA {\it Hubble Space Telescope}, or obtained
from the data archive at the Space Telescope Institute.

\end{document}